\documentclass[fleqn,usenatbib,useAMS]{mnras}

\usepackage{graphicx}	
\usepackage{amsmath}	
\usepackage{amssymb}	
\usepackage{multicol}   
\usepackage{bm}	     	
\usepackage{pdflscape}	
\usepackage[utf8]{inputenc}
\usepackage{array}
\usepackage{booktabs}

\newcommand{\kms}{\,km\,s$^{-1}$} 

\usepackage[T1]{fontenc}
\usepackage{ae,aecompl}

\usepackage{newtxtext,newtxmath}

\title[Synthetic observations of prestellar cores]{Synthetic observations of deuterated molecules in massive prestellar cores}

\author[Joaquin Zamponi et al.]{
        J. Zamponi$^{1,2}$\thanks{Contact e-mail: \href{mailto:jzamponi@mpe.mpg.de}{jzamponi@mpe.mpg.de}},
        A. Giannetti$^{3}$,
        S. Bovino$^{1}$,
        G. Sabatini$^{1,3,4}$,
        D. R. G. Schleicher$^{1}$,
        \newauthor
        B. K\"ortgen$^{5}$,
        S. Reissl$^{6}$ 
        and
        S. Wolf$^{7}$ 
        \\
$^{1}$Departamento de Astronom\'ia, Universidad de Concepci\'on, Esteban Iturra s/n Barrio universitario, Casilla 160, Concepci\'on, Chile\\
$^{2}$Max-Planck-Institut f\"ur Extraterrestrische Physik (MPE), D85748 Garching, Germany\\
$^{3}$INAF - Istituto di Radioastronomia - Italian node of the ALMA Regional Centre (It-ARC), Via Gobetti 101, I-40129 Bologna, Italy\\
$^{4}$Dipartimento di Fisica e Astronomia, Universitá degli Studi di Bologna, Via Gobetti 93/2, I-40129 Bologna, Italy\\
$^{5}$Hamburger Sternwarte, Universität Hamburg, Gojenbergsweg 112, D-21029 Hamburg, Germany\\
$^{6}$Heidelberg University, Center for Astronomy, Institute of Theoretical Astrophysics, Albert-Ueberle-Str. 2, 69120 Heidelberg, Germany\\
$^{7}$Institut für Theoretische Physik und Astrophysik, Christian-Albrechts-Universität zu Kiel, Leibnizstr. 15, 24118 Kiel, Germany\\
}

\date{Accepted XXX. Received YYY; in original form ZZZ}

\pubyear{2020}

\begin{document}
\label{firstpage}
\pagerange{\pageref{firstpage}--\pageref{lastpage}}
\maketitle

\begin{abstract}
 Young massive stars are usually found embedded in dense massive molecular clumps and are known for being highly obscured and distant.
 During their formation process, deuteration is regarded as a potentially good indicator of the very early formation stages.
 In this work, we test the observability of the ground-state transitions of ortho-H$_2$D$^+$ $J_{\rm {K_a, K_c}} = 1_{10}$-$1_{11} $, by performing interferometric and single-dish synthetic observations using magneto-hydrodynamic simulations of high-mass collapsing molecular cores, including deuteration chemistry. 
 We studied different evolutionary times and source distances (from 1 to 7~kpc) to estimate the information loss when comparing
 the column densities inferred from the synthetic observations to the column densities in the model. 
 We mimicked single-dish observations considering an APEX-like beam and interferometric observations using CASA and assuming the most compact configuration for the ALMA antennas. 
We found that, for centrally concentrated density distributions, the column densities are underestimated by about 51\% in the case of high-resolution ALMA observations ($\leqslant$1") and up to 90\% for APEX observations (17"). 
 Interferometers retrieve values closer to the real ones, however, their finite spatial sampling results in the loss of contribution from large-scale structures due to the lack of short baselines.
 We conclude that, the emission of o-H$_2$D$^+$ in distant massive dense cores is faint and would require from $\sim$1 to $\sim$7 hours of observation at distances of 1 and 7 kpc, respectively, to achieve a 14$\sigma$ detection in the best case scenario. 
 Additionally, the column densities derived from such observations will certainly be affected by beam dilution in the case of single-dishes and spatial filtering in the case of interferometers. 

\end{abstract}






\section{Introduction}
\label{sec:Introduction}
Massive stars, with masses above $\sim$8~M$_{\odot}$, significantly impact the environment in which they are born. They affect the thermal properties and chemical composition of the parent cloud via photoionization and dust heating onto the circumstellar discs of their neighbouring low-mass stars. They end their lives as supernovae, impacting the surroundings and the subsequent star formation by depositing high amounts of heavy elements and increasing the level of turbulence within the cloud.
Currently, a global consensus does not exist on how high-mass stars are formed, and attempting to clarify this question (e.g., ~\citealt{McKeeAndOstriker,ZinneckerAndYorke07,MotteBontemsAndLouvet} for a reviews), \citet{Bonnell97,Bonnell01} proposed the competitive core accretion model where all bound objects accrete gas from their surroundings; objects placed in the center of the cloud become (on average) more massive than in the outskirts thanks to the favourable conditions for accretion at the bottom of the potential well. 
\citet{McKeeAndTan02,McKeeAndTan03} proposed a scaled-up version of the low-mass star paradigm \citep{Shu87}, termed turbulent core accretion model, where prestellar cores are supposed to have high gas pressures, supersonic turbulence and significant magnetic support, leading to a rather slow, almost monolithic and unfragmented collapse.
One of the largest differences between these models is the predicted collapse timescale. Therefore, a proper analysis of such timescales is needed to distinguish between the formation scenarios.
A useful observational tool to measure such timescales are chemical clocks, i.e. tracers which show drastic abundance changes with density and temperature variations.
It is possible to combine the information from distinct molecules, but the possibility to use the abundance of a species in this way depends on the observability of each molecule transitions. 

Infrared Dark Clouds (IRDC), massive quiescent clouds that constitute a subset of the most likely birthplaces of the next generation of high-mass stars in our galaxy, are characterized by low gas temperatures ($T_{\rm gas}<$20K), high gas column densities ($N_{\rm gas}\sim10^{23-25}$ cm$^{-2}$) and a large degree of CO-depletion~(e.g., \citealt{CaselliAndCeccarelli12,Caselli13,Jorgensen20}; see also \citealt{BerginAndTafalla07} for a review). 
The absence of C-bearing molecules in the gas phase enables further chemical reactions to take place, such as the enhancement in abundance of deuterated molecules. 
Deuteration reactions start from the proton-deuteron exchange between HD and H$_3^+$, 
\begin{eqnarray}
{\rm H}_3^+ +{\rm HD}&\rightleftharpoons&{\rm H}_2{\rm D}^+ +{\rm H}_2 +\Delta E_1,\\
{\rm H}_2{\rm D}^+ +{\rm HD}&\rightleftharpoons&{\rm D}_2{\rm H}^+ +{\rm H}_2 +\Delta E_2,\\
{\rm D}_2{\rm H}^+ +{\rm HD}&\rightleftharpoons&{\rm D}_3^+ +{\rm H}_2 +\Delta E_3,
\end{eqnarray}
where $\Delta E_1$, $\Delta E_2$ and $\Delta E_3$ depend on the isomers involved in each reaction~\citep{Hugo07}.
In addition, H$_3^+$ and H$_2$D$^+$ can be destroyed by the following reactions
\begin{eqnarray}
{\rm H}_3^+ +{\rm CO}&\rightleftharpoons&\rm{HCO}^+ +{\rm H}_2,\\
{\rm H}_2{\rm D}^+ +{\rm CO}&\rightleftharpoons&{\rm DCO}^+ +{\rm H}_2, \label{eq:h2d+_to_dco+}
\end{eqnarray}
but since the gas is CO-depleted, they are quenched and the H$_2$D$^+$ 
abundance is expected to increase as the density of the gas increases~\citep{Dalgarno&Lepp84}.

Models of the evolution of such deuteration reactions \citep{Walmsley04,van_der_Tak05,Flower06,Sipila13}, show a maximum o-H$_2$D$^+$ abundance right before the formation of a protostellar object, which would then heat up the gas and evaporate the CO from the dust grain surfaces, converting H$_2$D$^+$ into DCO$^+$, according to reaction~(\ref{eq:h2d+_to_dco+}).

The different isomers of the H$_2$D$^+$ molecule represent also a very powerful proxy to estimate the ortho-to-para ratio of H$_2$ (e.g., \citealt{Flower06,Hugo09,Brunken14}), a highly uncertain but fundamental parameter to understand the deuteration process in star-forming regions.
Unfortunately, observations of low energy transitions from para-H$_2$D$^+$ suffer from large atmospheric attenuation at terahertz frequencies, hence, most of the efforts have been directed to observe the sub-millimeter transitions of ortho-H$_2$D$^+$ $J_{\rm {K_a, K_c}} = 1_{10}$-$1_{11}$ at 372.42~GHz.
Several detections have been reported toward low-mass \citep{Caselli03,Vastel06,Caselli08,Parise11,Friesen14,Brunken14,Miettinen20} and high-mass sources \citep{Harju06,Swift09,Pillai12,Giannetti19}.
The first survey of o-H$_2$D$^+$ in high-mass star-forming regions has been recently carried out, reporting 17 detections out of a sample of 106 sources~\citep{SabatiniSUB}.

Massive prestellar clumps are rather distant (>1~kpc; \citealt{Giannetti14} and \citealt{Konig17}), with the exception of a few cases \citep{ZinneckerAndYorke07}, and therefore detection of o-H$_2$D$^+$ is rather difficult, requiring high angular resolutions and long integration times.

In this work, we performed a study of the observability of the o-H$_2$D$^+$ $J_{\rm {K_a, K_c}} = 1_{10}$-$1_{11}$ molecular line using magneto-hydrodynamical (MHD) simulations of collapsing magnetized prestellar cores, performed by \citet{Kortgen17}, which included the evolution of deuteration reactions.
For this, we performed a set of radiative transfer (RT) simulations using these simulated cores as synthetic sources (see~\ref{subsec:Synthetic_sources}) and then post-processed them by adding instrument-related effects. 
In this last step, we distinguished between single-dish and interferometric observations, looking for an understanding of the key differences that may arise when observing the same source with both techniques. 
Finally, we derived column densities of o-H$_2$D$^+$ from the resulting intensity distributions and compared our results with values reported in the literature and to the physical column densities measured in the model. 

In section~\ref{sec:Methods} we describe the steps followed for each synthetic observation, in section~\ref{sec:Results} we show the radial distributions of the synthetic maps and the column densities derived, in section~\ref{sec:Discussion} we discuss the limiting cases of our results in terms of source distance and observing time and finally, we provide a summary and conclusions in section~\ref{sec:Summary_and_conclusion}.

\section{Methods}
\label{sec:Methods}
\subsection{Workflow}
\label{subsec:Workflow}
\begin{figure}
    \includegraphics[width=\columnwidth,trim=7cm 2.3cm 7cm 2.3cm, clip]{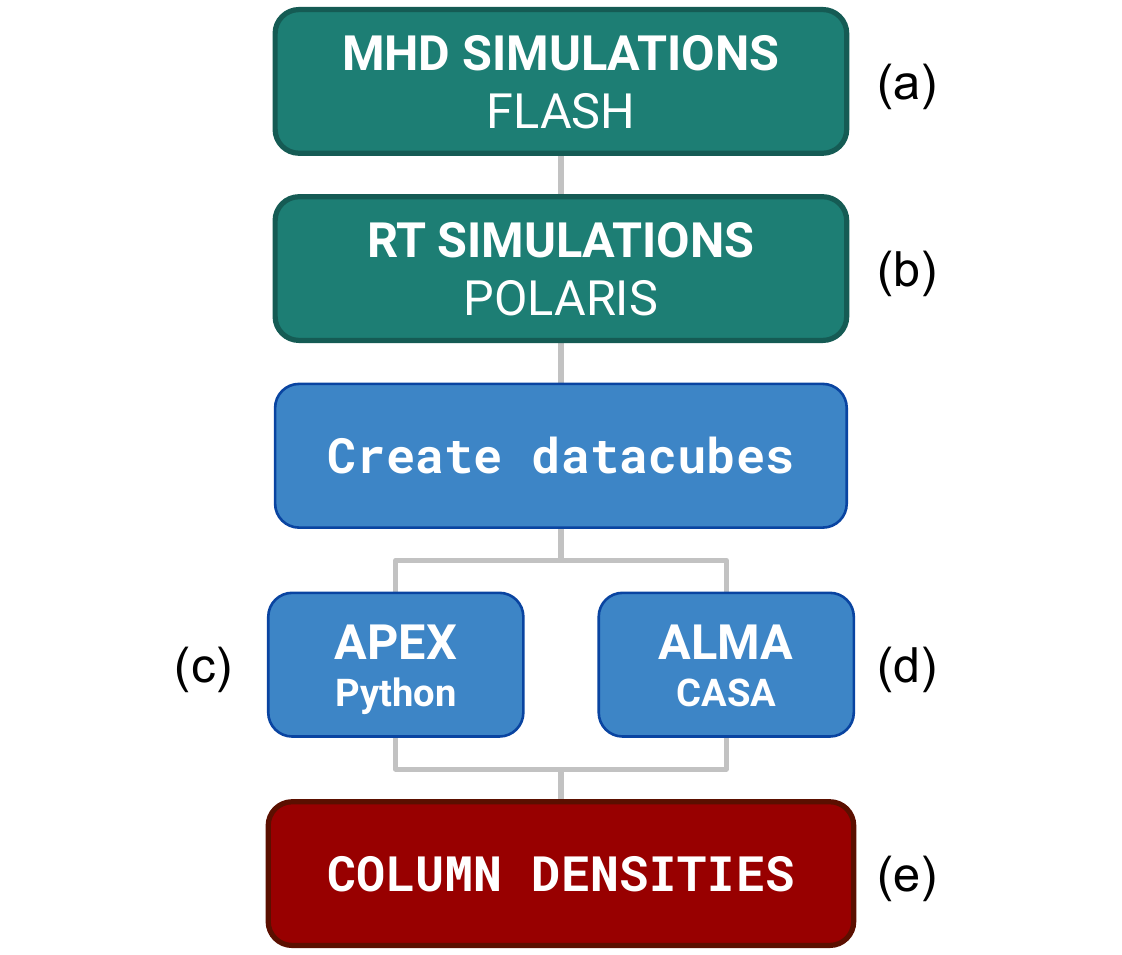}
    \caption{Workflow of each synthetic observation. The input MHD simulations are ray-traced using the POLARIS RT code and then post-processed to distinguish between interferometric and single-dish simulations, using the CASA software and a Python module written for this project, respectively. Each resulting intensity distribution is then converted into column densities for further comparison with real data.}
    \label{fig:workflow}
\end{figure}
To perform a successful synthetic observation, three main ingredients are needed: a synthetic source, a ray-tracer and a synthetic detector.
The synthetic source, or model, can be either a simple density and temperature distribution or a full three dimensional magneto-hydrodynamic (MHD) simulation~\citep{Haworth18} (Fig.~\ref{fig:workflow}a). 
The ray-tracing is done by an RT code, which calculates the propagation of light within the source, accounting for emission, absorption and scattering, and creates a resulting flux distribution attenuated at a given distance (Fig.~\ref{fig:workflow}b).
The synthetic detector is the numerical array in which this intensity distribution is stored.
On top of it, instrument related effects must be applied, such as convolution with the telescope beam (in the case of single-dish telescopes) and reconstructed beam (in the case of inteferometers), addition of noise and image reconstruction from a complex visibility, in the case of interferometers (Fig.~\ref{fig:workflow}c,d). 
Thus, in order to make synthetic observations as realistic as possible, the intensity maps from the RT simulations must be post-processed based on the specific properties of the telescope of interest.
The fluxes resulting from each synthetic observation are finally converted into column densities and the results between different observing setups compared (Fig.\ref{fig:workflow}e).
All the functions and routines used in steps (c), (d) and (e) from Fig.\ref{fig:workflow}, are provided in an online repository\footnote{\url{https://github.com/jzamponi/synthetic_module}}.

\subsection{Synthetic source}
\label{subsec:Synthetic_sources}
The synthetic source used in this work is an isolated magnetized massive prestellar core taken from the 3D ideal-MHD simulations performed by~\citet{Kortgen17}.
They employed the \textit{FLASH} code (v4.2.2)~\citep{FLASH}, coupled with the \textit{KROME} package~\citep{Grassi14} to follow the deuteration chemistry of light hydrides.

\subsubsection{Initial conditions}
\label{subsubsec:Initial_conditions}
The core is a Bonnor-Ebert (BE) sphere~(\citealt{Bonnor56} and \citealt{Ebert95}), initialized as supersonically turbulent and assumed to collapse isothermally at $T_{\rm gas}=15$~K. The initial conditions of the core are listed in Table~\ref{tab:initial_conditions}.
The maximum spatial resolution achieved during runtime was $\Delta x_{\rm max}=235$~AU. 
Sink particles were also included for gravitationally bound gas, with a density threshold of $n_{\rm sink}=3\times10^7$~cm$^{-3}$ and an accretion radius of $R_{\rm accr}=2\Delta x_{\rm max}=470$~AU. This is the radius within which the gas is considered to be absorbed by the dense object and is meant to represent the formation of a protostar.

\begin{table*}
    \caption[Initial parameters of the cores selected]{Initial parameters of the core selected from~\citet{Kortgen17}, labeled Lmu10M2. The name refers to a source within their sample that has a low surface density, a mass-to-flux ratio of 10 and a Mach number of 2.}
    \label{tab:initial_conditions}
    \centering
    \setlength\tabcolsep{6pt} 
    \begin{tabular}{l c c c c c c c c c }
    \toprule[0.6pt]
     & Surface & Core & Core & Av. field & Mass-to- & Mach & Virial & Number of & Free-fall \\
    Run & density  & radius & mass & strength & flux ratio & number & parameter & Jeans & time \\
      & (g cm$^{-2}$) & (pc) & (M$_{\odot}$) & ($\mu$G) & $\mu/\mu_{\rm crit}$ & $\mathcal{M}_{\rm turb}$ & $\alpha_{\rm vir}$ & masses & (kyr)\\ 
    \midrule[0.3pt]
    Lmu10M2 & 0.14 & 0.17 & 60 & 27 & 10 & 2 & 0.64 & 24 & 149 \\
    \bottomrule[0.6pt]
    \end{tabular}
\end{table*}

\begin{figure*}
    \hspace{-0.05\textwidth}
    {\Large 0.1~$t_{\rm ff}$ \hspace{0.20\textwidth} 0.5~$t_{\rm ff}$ \hspace{0.20\textwidth} 1~$t_{\rm ff}$}\\
    \includegraphics[trim=0.5cm 0.1cm 0.7cm 0.5cm,clip, width=17cm]{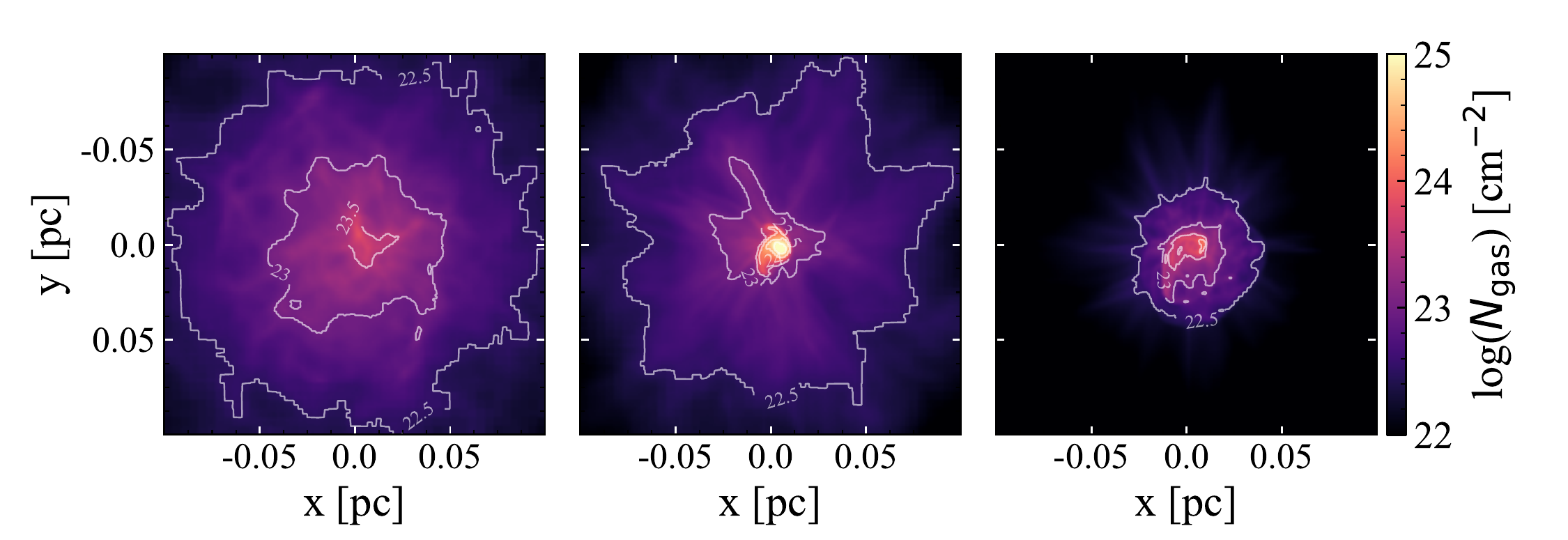}
    \caption{Snapshots of the time evolution at 0.1, 0.5 and 1~$t_{\rm ff}$ for the synthetic core Lmu10M2. White contours are the gas column density at 10$^{22}$, 10$^{23}$, 10$^{24}$ and 10$^{25}$cm$^{-2}$.}
    \label{fig:mhd_snapshots}
\end{figure*}

\subsubsection{Deuteration chemistry}
\label{subsubsec:Deuteration_chemistry}
The simulations from~\citet{Kortgen17} included for the first time an accurate non-equilibrium chemical network, with 21 chemical species in the gas phase and dust grains, along with their ionized states. 
The network solved the deuteration reactions based on~\citet{Walmsley04}, who assumed total depletion of elements heavier than He, which accounts for the CO freeze-out onto dust ice mantles that has been observed to be effective at the densities studied here ($\gtrsim10^4$~cm$^{-3}$; \citealt{Caselli99, Tafalla02, Giannetti14}). This assumption allows to significantly reduce the network and the computational time of the simulation, while still providing realistic results~(e.g. \citealt{Sabatini19} and \citealt{Bovino19}). 
Electron attachment, recombination of positive ions and grain surface reactions were also included, such as the formation of H$_2$ and HD, with the exception of D$_2$, which is mainly formed in the gas-phase.
The full chemical network was solved and evolved using the \textsc{KROME} package \citep{Grassi14}.

\subsection{Radiative transfer simulations}
\label{subsec:Radiative_transfer_simulations}
We performed ray-tracing RT simulations with the code \textit{POLARIS}\footnote{\url{http://www1.astrophysik.uni-kiel.de/~polaris}} (v4.06)~(\citealt{Reissl16} and \citealt{Brauer17}) based on the temperature and density distribution of ten timesteps from the collapsing core simulations, starting from slightly after the zero-age (0.1~$t_{\rm ff}$) collapse until 1~$t_{\rm ff}$ (one free-fall time), in intervals of 0.1~$t_{\rm ff}$ (see Fig.~\ref{fig:mhd_snapshots}).
We simulated the emission of the o-H$_2$D$^+$ $J_{\rm {K_a, K_c}} = 1_{10}$-$1_{11}$ transition at 372.42~GHz, which is optically thin in all of our simulations. 
Dust was not included in our setup.
The setup included distances from 1 up to 7 kpc, in intervals of 1~kpc.
For each time and distance, we have created square maps of 512$\times$512 pixels, where each pixel has a physical scale of $\sim$80x80~AU (i.e., each side of the map corresponds to 0.2~pc).
The spectral range covers 10~km s$^{-1}$ centered on the rest frequency of our line of interest, i.e. 372.42~GHz.
We fixed the spectral resolution of each datacube to 0.03~km\,s$^{-1}$ ($\sim$38~kHz) for all simulations by splitting the spectral range into 333 channel maps, to represent the highest resolution offered by the FLASH$^+$ dual-frequency MPIfR principal investigator (PI) receiver~\citep{FLASH+} at the frequency of 372.42~GHz, mounted at the \textit{Atacama Pathfinder EXperiment Telescope} (APEX;~\citealt{Gusten06}).  
Based on the gas column density distribution of the synthetic sources (see Fig.~\ref{fig:mhd_snapshots}), the lowest number densities are around 1.62$\times$10$^5$~cm$^{-3}$, for a core radius of 0.2~pc.
Such number densities are above the critical density of the o-H$_2$D$^+$ line emission at 372.42~GHz  (1.3$\times$10$^5$~cm$^{-3}$; \citealt{Hugo09}) and therefore Local Thermodynamic Equilibrium (LTE) can be assumed when computing the level populations.  
The assumption of LTE means that the excitation temperature $T_{\rm ex}$ and the kinetic gas temperature $T_{\rm gas}$ were equal (15~K), and then 
\begin{equation}
    T_{\rm gas}=T_{\rm ex}=\frac{h\nu_{ij}}{k_{\rm B}}\left[\ln{\left(\frac{g_i\,n_j}{g_j\,n_i}\right)}\right]^{-1},
\end{equation}
with $g_i$ and $g_j$ the statistical weights, and  $n_i$ and $n_j$ the level populations following the Boltzmann distribution.
The molecular data of o-H$_2$D$^+$ used in our simulations were obtained from the Leiden Atomic and Molecular Database\footnote[2]{\url{http://home.strw.leidenuniv.nl/~moldata}} (LAMDA;~\citealt{Schoier05}).

\subsection{APEX simulations}
\label{subsec:APEX_simulations}
The single-dish synthetic observations are done in image post-processing. We first converted the output from POLARIS into datacubes, and then proceeded to add instrument related effects, i.e. convolution of the image with a telescope beam and the addition of noise (Fig.~\ref{fig:workflow}c).
All the functions used for this post-processing are available in the online repository mentioned in section~\ref{subsec:Workflow}.

\subsubsection{PSF convolution}
We convolved the datacubes with a 2D Gaussian kernel, resembling the Point-Spread-Function (PSF) of a parabolic single-dish telescope.
We used a Full-Width-at-Half-Maximum (FWHM) of 16.8 arcseconds, that is the effective resolution achieved by the APEX 12-m dish at a frequency of 372.4~GHz. 
After the convolution, we converted the maps from Jy\,pixel$^{-1}$ into Jy\,beam$^{-1}$ by rescaling the flux with the ratio of the area of a gaussian beam over the area of a square pixel, as
\begin{equation}
  \frac{F}{{\rm Jy\,beam}^{-1}} = \frac{\pi}{4\ln 2}\frac{\theta_{\rm maj}\,\theta_{\rm min}}{{\rm arcsec}^2}\left(\frac{\rm pix\,size}{\rm arcsec}\right)^{-2} \frac{F}{{\rm Jy\,pixel}^{-1}},
\end{equation}
where $\theta_{\rm maj}$ and $\theta_{\rm min}$ are the FWHM of the major and minor axis of the beam in arcseconds, here assumed to be equal.
At a distance of 1~kpc, the APEX-beam corresponds to 16800~AU.
This means that these observations were spatially unresolved almost during the entire evolution.

Convolution onto the images was performed using the \texttt{convolve\_fft}\footnote{\url{https://docs.astropy.org/en/stable/convolution}} function from the Astropy Python package~\citep{astropy2013,astropy2018}.
This function performs a Fourier-space convolution of 2D data-matrix with a Gaussian kernel of standard deviation $\sigma={\rm FWHM}\,/\,\sqrt{8\ln 2\,}$, where FWHM is the angular resolution in number of pixels.

\subsubsection{Addition of noise}
We converted the images from Jy beam$^{-1}$ to brightness temperature ($T_{\rm b}$) and added Gaussian noise to them, with a sigma ($T_{\rm rms}$) derived from the equation 
\begin{equation}
    \label{eq:T_rms_single-dish}
    T_{\rm rms} = \frac{T_{\rm sys}}{\sqrt{\Delta\nu\,t_{\rm int}}}, 
\end{equation}
where $\Delta\nu$ is the spectral resolution, $t_{\rm int}$ is the integration time and $T_{\rm sys}$ is the system temperature~\citep{Kraus_Radio_Astronomy}.
To obtain realistic values of $T_{\rm sys}$ for a given observing setup, we used the APEX Observing time calculator\footnote{\url{http://www.apex-telescope.org/heterodyne/calculator}}.
The noise level added to the simulations was $T_{\rm rms}\sim8.6$~mK, corresponding to $T_{\rm sys}\sim543$~K for an integration time of 6~hrs, a source elevation of 45~deg, a precipitable-water-vapor (pwv) level of 0.5 (e.g. \citealt{Miettinen20}) and a spectral resolution of $\Delta v=0.3$~km\,s$^{-1}$, achieved by binning the spectra with a factor of 10.

\subsection{ALMA simulations}
\label{subsec:ALMA_simulations}
The fourth step of each observation was to simulate the \textit{Atacama Large Millimeter/submillimeter Array} (ALMA;~\citealt{Wootten09}) response (Fig.~\ref{fig:workflow}d). 
We simulated the interferometric observations by using the 
\textit{Common Astronomy Software Applications package} (CASA) tasks SIMOBSERVE\footnote{\url{https://casa.nrao.edu/docs/taskref/simobserve-task.html}} and TCLEAN\footnote{\url{https://casa.nrao.edu/docs/taskref/tclean-task.html}}.
SIMOBSERVE is used to create a visibility measurement set from an image model and TCLEAN to reconstruct an image out of the visibility table using the CLEAN algorithm.
An example script for the ALMA simulations is provided in the online repository mentioned in section~\ref{fig:workflow}.
The image models used as input were the output from the radiative transfer simulation, in units of Jy\,pixel$^{-1}$. 
Given the frequency of the transition observed, we used the most compact array configuration C43-1 in band 7 (Cycle 7), obtaining a synthesized beam of about $1".01\times0".92$, selected to achieve high sensitivity since the o-H$_2$D$^+$ emission is rather faint in these distant sources.
This configuration also aids in limiting the spatial filtering of the extended emission.
Based on the same argument, we cleaned with natural weighting for all images, aiming for improved sensitivity over angular resolution, since all sources are resolved in the interferometric case.
The cleaning was performed using an image size of 512 pixels per side and a cell size of $\sim$0.08" for the simulations at 1~kpc. For larger distances, the cell size was scaled accordingly to maintain the coverage of 80~AU per pixel and 512~pixels per side.
For the noisy simulations we cleaned interactively, to define a proper mask for source at each timestep and distance, and increased the number of iterations from 500 up to 15000, checking the residual map at each major cycle. The cleaning threshold was also adjusted interactively and a Hogbom deconvolver was used \citep{Hogbom74}. 
The gridder was set to \textit{standard} because the noisy simulations consisted of one single pointing, aimed at the core center, while the noiseless simulations used a \textit{mosaic} gridder to cover the whole map with multiple pointings. 
This strategy for the noisy simulation was employed to achieve a higher Signal-to-Noise (SNR) for a certain integration time, which would have been split among the multiple pointings in the case of a mosaic. For the noiseless simulations the SNR was not an issue and then we cover the whole model image.
Additionally, the slight differences in frequency across the line emission results in slightly different synthesized beams. To keep it constant along the line, we set the restoring beam to \textit{common}.
The spectral channel width was 0.03~\kms, although we also binned the spectrum up to 0.3 and 1~\kms when analyzing the low-SNR observations at large source distances. 
The synthetic sources were given the sky position 19h00m00s -40d00m00s (J2000) and they reached a maximum elevation of $\sim$70~degrees from the horizon after 7.5~hrs. 
Thermal noise was added by the \texttt{simobserve} task, based on a sky temperature of 114.3~K, atmospheric zenit opacity ($\tau_0$) of 0.579, pwv of 0.658 and a ground temperature of 269~K.
This led to noise levels of $\sim$1.4~mJy beam$^{-1}$ (9~mK) for a 10~hrs integration time.

Part of the analysis included simulations using the \textit{Atacama Compact Array} (ACA;~\citealt{Iguchi09}). 
ACA is used to patch the shorter baselines in the \textit{uv}-plane, not covered by the ALMA main array, retrieving information from large scale emission.

\subsection{Derivation of column densities}
\label{subsec:Derivation_of_column_densities}
To compare our column densities derived from synthetic observations ($N_{\rm APEX}$ and $N_{\rm ALMA}$) to the model values ($N_{\rm model}$), we integrated the o-H$_2$D$^+$ number density within the simulation box along the light-of-sight (LOS) and then derived $N_{\rm model}$ by averaging the column densities over the area subtended by the beam of interest.
Column densities from the synthetic observations were derived from the flux, averaged over the telescope beam and weighted by the beam response.  
For this we followed the procedure described in \citet{MagnumAndShirley}, similarly used in \citet{Vastel06, Caselli03, Busquet10} and \citet{Parise11}, where the total column density ($N$) of a given species X is given by
\begin{equation}
    \label{eq:column_density}
    N(X) = \frac{8\pi\nu^3}{c^3}\frac{Q(T_{\rm ex})}{g_{u}A_{ul}}\frac{e^{E_{u}/T_{\rm ex}}}{e^{h\nu/k T_{\rm ex}}-1}\int\tau dv,
\end{equation}
where $u$ and $l$ refer to the upper and lower levels of the transition, respectively,
$k$ and $h$ are the Boltzmann and Planck constants, respectively, $\nu=372.42$~GHz is the frequency of the transition $J_{\rm {K_a, K_c}} = 1_{10}$-$1_{11}$. The statistical weight of the upper level and the Einstein coefficient for the transition are $g_u=9$ 
and $A_{ul}=1.08\cdot10^{-4}$~s$^{-1}$, respectively. 
$E_u=17.87$~K is the energy of the upper level, $Q(T_{\rm ex})=10.3375$ is the partition function of the molecule for $T_{\rm ex}=15$~K and $\tau$ is the optical depth of the line.
The derivation of the last two parameters is described in the following sections.

\subsubsection{The Partition function}
\label{subsubsec:The_partition_function}
To calculate the partition function, we followed the description given by \citet{MagnumAndShirley}:
the partition function of a molecule in a gaseous state is a function of its electronic ($Q_e$), vibrational ($Q_v$), rotational ($Q_r$) and nuclear ($Q_n$) partition functions. 
For most interstellar molecules their ground electronic state is stable, leading to $Q_e$=1. 
Also, considering molecules that are in their ground vibrational state ($Q_v$=1), then, the total partition function is reduced to 
\begin{eqnarray*}
Q_{\rm tot}&=&Q_e\,Q_v\,Q_r\,Q_n,\\
Q_{\rm tot}&=&Q_r\,Q_n,\\
Q_{\rm tot}&=&\sum_i g_i\exp\left(-\frac{E_i}{k T_i}\right),
\end{eqnarray*}
with $E_i$ and $T_i$ the energy and temperature of the level $i$, $g_i=g_u=g_Kg_Ig_J$, where $g_K$ is the degeneracy associated with the internal quantum number $K$ in asymmetric molecules, due to projections of the total angular momentum onto a molecular axis, $g_I$ is the nuclear degeneracy, set to 3 for o-H$_2$D$^+$, and $g_J$ is the rotational degeneracy, also set to 3 for o-H$_2$D$^+$ as $J=1$ for the transition $J_{\rm {K_a, K_c}} = 1_{10}$-$1_{11}$. 
For symmetric and slightly asymmetric molecules, $g_K=1$, $g_J=(2J+1)=3$ and $g_I=3$. Therefore, the total partition function is $Q(T_{\rm ex}=15~K)=10.3375$.

\subsubsection{The optical depth}
\label{subsubsec:The_optical_depth}
We start from the source brightness temperature ($T_{\rm b}$) which can be obtained from the equation
\begin{equation}
    \label{eq:brightness_temperature}
    T_{\rm b} = [J_{\nu}(T_{\rm ex})-J_{\nu}(T_{\rm bg})](1-e^{-\tau})\,,
\end{equation}
with $T_{\rm ex}=15$~K, $T_{\rm bg}$ the background temperature, given by the Cosmic Microwave Background (CMB) temperature of 2.7~K, and 
\begin{equation}
    \label{eq:Radiation_temperature}
    J_{\nu}=\left(\frac{h\nu}{k_B}\right)\left(e^{\frac{h\nu}{k_BT}}-1\right)^{-1}.
\end{equation}
To solve for the optical depth ($\tau$) from equation~(\ref{eq:brightness_temperature}), we use
\begin{equation}
    \label{eq:Intensity_to_Tb}
    T_{\rm b}=\frac{\lambda^2}{2k_{\rm B}\Omega}S,
\end{equation}
with $\lambda$ the wavelength of observation and $S$ the flux density over a certain solid angle $\Omega$, or from the intensity distribution $I$
\begin{equation}
    \label{eq:Intensity_to_Tb_rescaled}
    T_{\rm b}=1.222\cdot10^3\frac{I}{\nu^2\theta_{\rm min}\theta_{\rm maj}},
\end{equation}
with $I$ in mJy\,beam$^{-1}$, $\nu$ the frequency  of observation in GHz and $\theta_{\rm min,maj}$ the minor and major axis of the telescope beam (synthesized beam for interferometers) in arcseconds.
The observed $T_{\rm b}$ induces an antenna temperature in the receiver, which must be converted into a main-beam brightness temperature $T_{\rm mb}$ by correcting with the main-beam efficiency ($\eta_{\rm mb}$) of the telescope. 
For APEX, at 372~GHz, this value is $\eta_{\rm mb}=0.69$ and for ALMA, the conversion was performed by CASA during runtime. 
Then, we define the optical depth as
\begin{eqnarray}
e^{-\tau}&=&1-\frac{T_{\rm mb}}{J_{\nu}(T_{\rm ex})-J_{\nu}(T_{\rm bg})},\\
    \tau &=&-\ln{\left(1-\frac{T_{\rm mb}}{J_{\nu}(T_{\rm ex})-J_{\nu}(T_{\rm bg})}\right)}\label{eq:optical_depth},
\end{eqnarray}
which is now based only on the value of $T_{\rm ex}$ and the received brightness temperature.
This equation shows the importance of a properly chosen excitation temperature when deriving column densities from observations \citep{Caselli03,Vastel04}.
In our case, the assumption of LTE for the calculation of the level populations constrain this free parameter to be $T_{\rm ex}=T_{\rm gas}$.

For the comparison of observed quantities with the model values, in the following sections we derive column densities over several angular scales, depending on the telescope of interest.
When computing column density ratios, the same angular scales were used for the model and synthetic values.

\section{Results}
\label{sec:Results}

\subsection{Radial distributions}
\label{subsec:Radial_distributions}
We first studied the fluxes, intensities and column densities when derived from the output of the radiative transfer simulation, the ALMA simulation and also ALMA combined with ACA.
None of these simulations includes thermal noise, in order to avoid any biased interpretation due to low signal-to-noise ratios and also to show that these effects are intrinsic independent of noise.
For this analysis we considered three timesteps, chosen to match those depicted in Fig.~\ref{fig:mhd_snapshots}. 
The results of the analysis, placing the source at a 1~kpc distance are shown in Fig.~\ref{fig:radial_distributions_1kpc}.
The three columns represent the stage at 16, 74 and 149~kyr, from left to right, respectively. Fluxes are shown in the top row, intensities in the middle row and column densities in the bottom row. Each panel shows the results from the radiative transfer simulations in solid-green, from ALMA simulations in dashed-orange and from ALMA combined with ACA in dotted-blue. In the bottom row, we included the column densities obtained from the MHD simulation for reference, shown in semi-dashed-red.
Since they are all beam-averaged quantities, we plot values averaged over concentric circles of radii from 0.1 to 22 arcseconds, centered on the center of the image. 
All images at 1~kpc have a side length of 40~arcseconds.

\subsubsection{Flux distribution}
\label{subsubsec:Flux distribution}
The radial flux distribution is an increasing function of the radius because of the larger area over which it is integrated.
The increase is rather smooth for the early stage, at 16~kyr, since the gas surface density distribution is also a smooth function of radius, slightly evolved from the initial Bonnor-Ebert density distribution. 
Departures from a constant increase are seen in the later stages, most likely explained by the centrally concentrated gas.
The interferometric simulations show large differences relative to the radiative transfer case.
Such differences are also larger for larger radii.
A possible explanation for this could be the fact that the Largest Angular Scale (LAS) of ALMA at the compact configuration C-43 (Band 7) is 8.45 arcseconds in diameter, while the size of the map is 41 arcseconds (at 1~kpc), therefore, part of the extended flux is not detected by ALMA.
It is also appreciated how the inclusion of ACA in the observation, allows to recover emission from larger angular scales, increasing the total flux and making them closer to those from the radiative transfer simulation.

\subsubsection{Intensity distribution}
\label{subsubsec:Intensity distribution}
Large differences between the interferometric simulations and the radiative transfer are seen, although, smaller toward the later stages.
This could be explained by the fact that most of the dense gas is found within the largest angular scale from 74~kyr and on, and therefore the effect of spatial filtering is reduced. 

\subsubsection{Column density distribution}
\label{subsubsec:Column density distribution}
\begin{figure*}
    \centering
    \includegraphics[trim=2cm 2cm 2cm 2cm,width=1.0\textwidth]{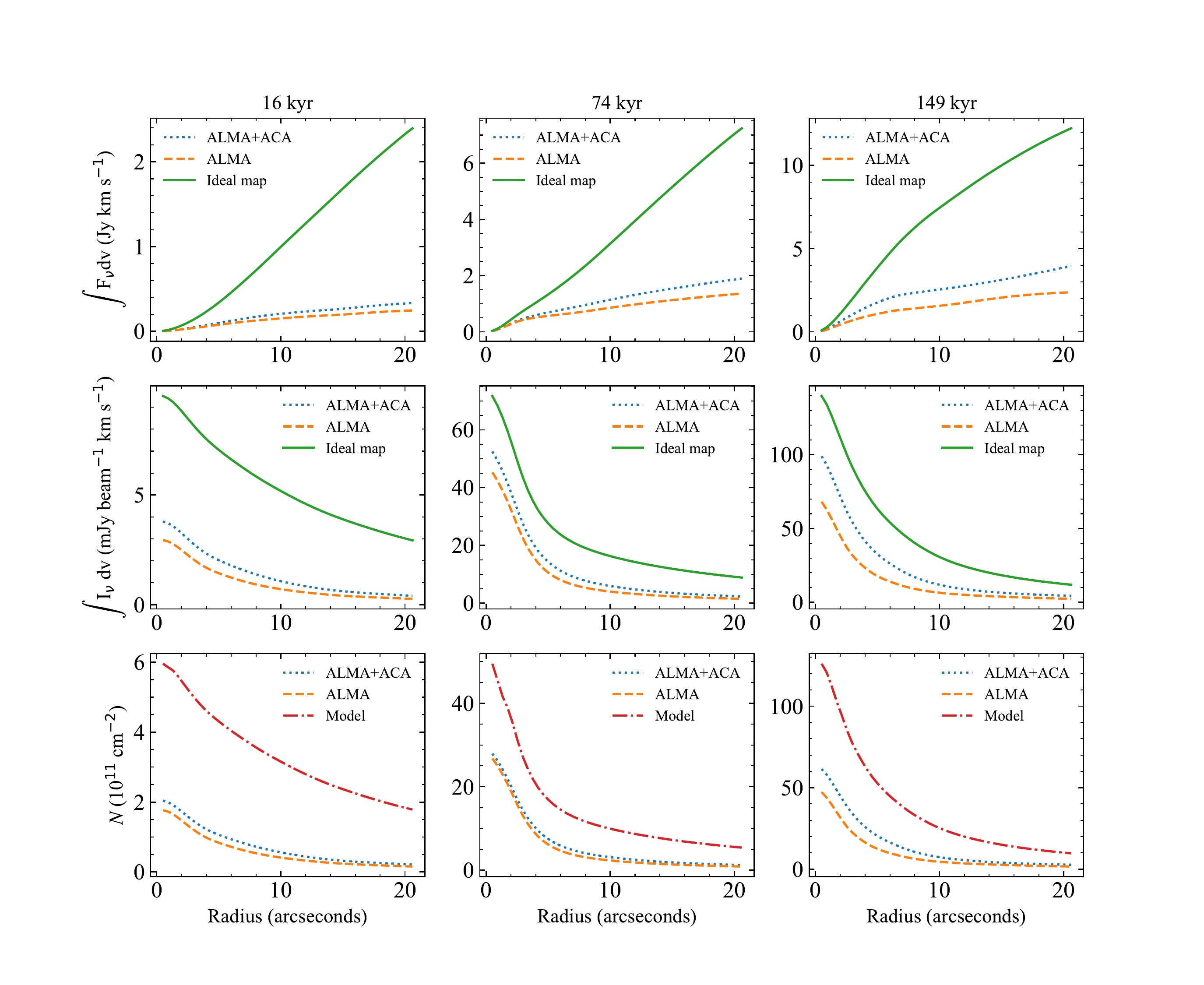}
    \caption{Radial distribution of the flux (top row), intensity (middle row) and column density (bottom row) at 1kpc, averaged over concentric circles of radii from 0.1 to 20 arcseconds, centered on the center of the image. The distributions are shown for three evolutionary stages, at 16 kyr (0.1$t_{\rm ff}$),  74 kyr (0.5$t_{\rm ff}$; before sink formation) and 149 kyr (1$t_{\rm ff}$), same as those depicted in Fig.~\ref{fig:mhd_snapshots}. The fluxes and intensities are shown for the ouput of the radiative transfer simulations (Ideal map) and for ALMA and ALMA combined with ACA. In the case of column densities, the model curve represent the physical column density integrated along the simulation box.}
    \label{fig:radial_distributions_1kpc}
\end{figure*}

\begin{figure*}
    \centering
    \includegraphics[trim=2cm 2cm 2cm 2cm,width=1.0\textwidth]{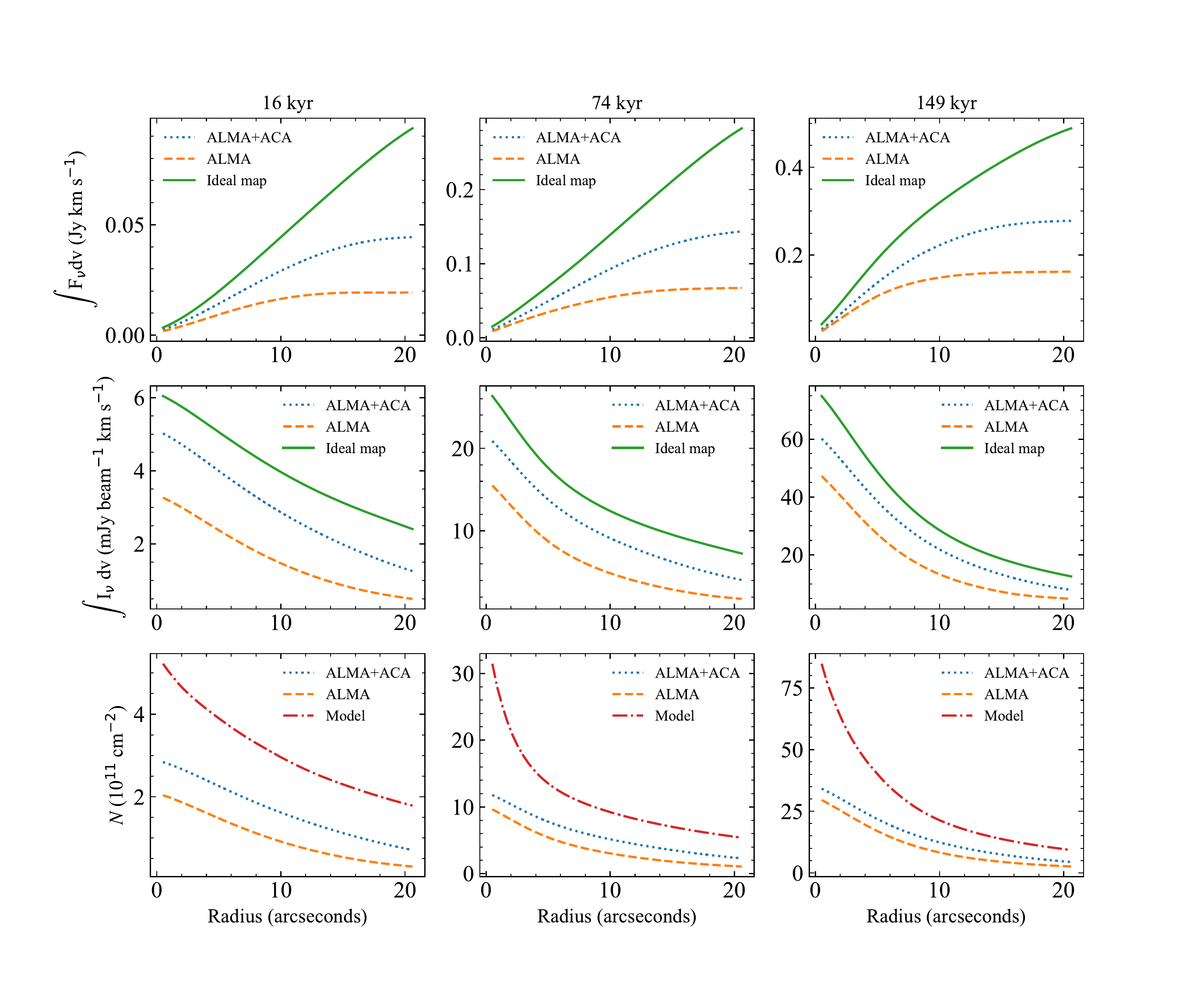}
    \caption{Same as figure~\ref{fig:radial_distributions_1kpc} but for a 5~kpc distance.}
    \label{fig:radial_distributions_5kpc}
\end{figure*}

\begin{figure*}
    \centering
    \includegraphics[trim=2cm 2cm 2cm 2cm,width=1.0\textwidth]{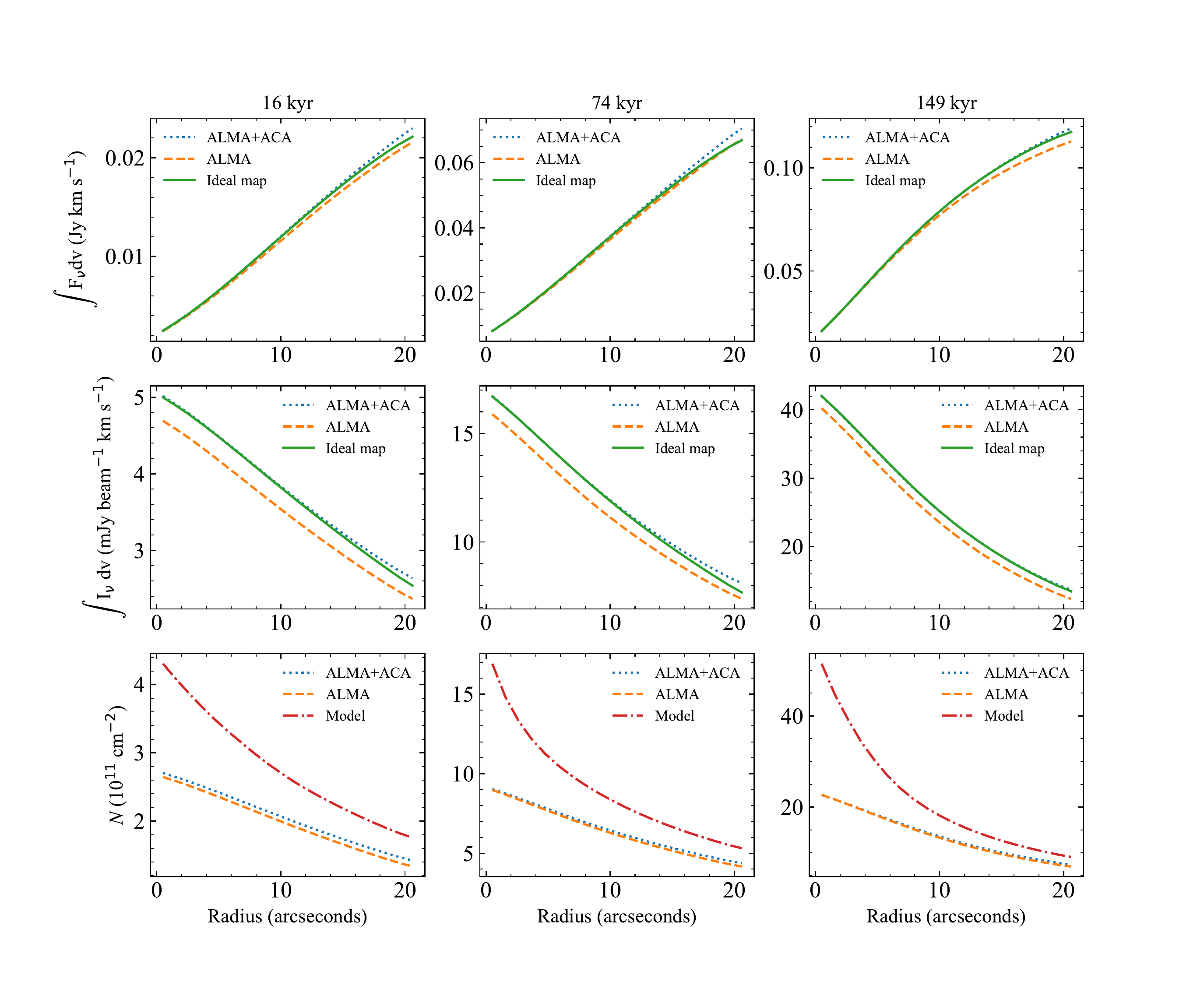}
    \caption{Same as figure~\ref{fig:radial_distributions_1kpc} but for a 10~kpc distance.}
    \label{fig:radial_distributions_10kpc}
\end{figure*}

The column density distribution is derived from the intensity using equation~(\ref{eq:column_density}) and therefore its radial distribution is similar to that of intensity. 
For the column density analysis, we included the values derived from the MHD simulations, labeled $N_{\rm model}$, in order to quantify the decrease in the measurement of column densities relative to the output of the ALMA simulations. 
The model values were obtained by integrating the o-H$_2$D$^+$ number density along the LOS and averaged over concentric circles.
The radii of such circles are the corresponding physical scales of apertures from 1 to $\sim$22 arcseconds, at a distance of 1~kpc (see Fig.~\ref{fig:radial_distributions_1kpc}), 5~kpc (see Fig.~\ref{fig:radial_distributions_5kpc}) and 10~kpc (see Fig.~\ref{fig:radial_distributions_10kpc}).
On the other hand, the observationally-derived (ALMA and ALMA plus ACA) column densities were obtained by ray-tracing the model number densities to obtain datacubes of intensity distributions.
These datacubes were used to generate complex visibilities and then cleaned, as described in section~\ref{subsec:ALMA_simulations}. 
These cubes where then converted into optical depths (following equation~\ref{eq:optical_depth}) and integrated along the frequency axis to obtain column densities (using equation~\ref{eq:column_density}).
Our main purpose here is to check how well the physical quantity can be recovered from the synthetic observation.

The differences between ALMA and the model are as high as in the case for the intensity and follow the same trend for later timesteps.
One possibility to explain such a large difference, is the spatial filtering produced by the incomplete spatial sampling of interferometers.
In order to test this idea, we reproduced a similar analysis using a source distance of 5 and 10~kpc, shown in Fig.~\ref{fig:radial_distributions_5kpc} and~\ref{fig:radial_distributions_10kpc}.
For the case at 5~kpc, the difference between the model and ALMA are clearly smaller, and even more so when including ACA. 
The case at 10~kpc shows an indiscernible difference between the RT and ALMA+ACA simulations, especially for the later stages. 

\section{Discussion}
\label{sec:Discussion}
\subsection{Comparison of inferred o-H$_2$D$^+$ column densities from ALMA and APEX}
\label{subsec:Comparison_of_column_densities}

\begin{figure}
    \centering
        \begin{center}
            \textbf{\Large Noiseless}
        \end{center}
    \includegraphics[width=1.0\columnwidth, trim=1cm 0 1cm 0]{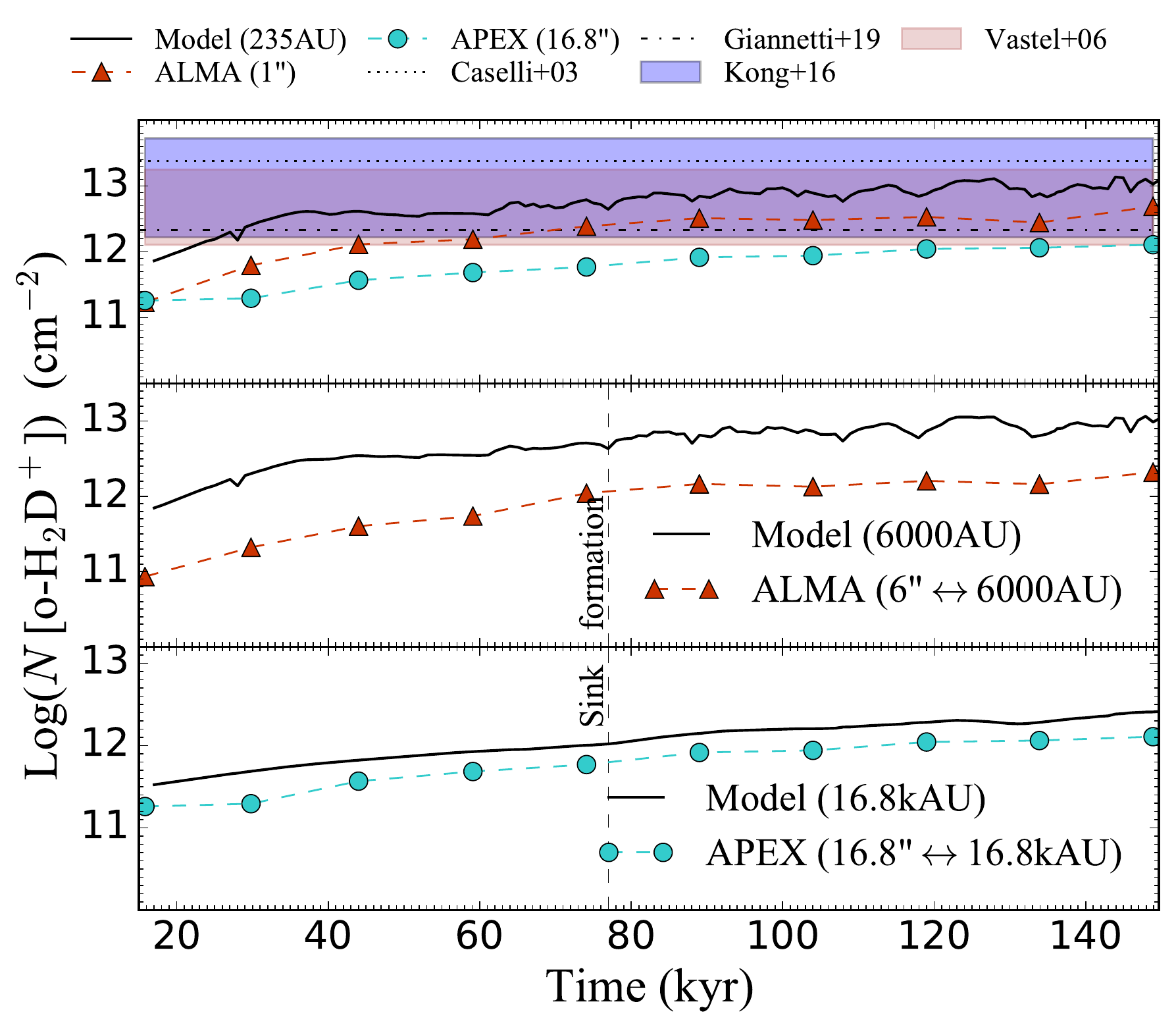}\\
        \begin{center}
            \textbf{\Large Noisy}
        \end{center} 
    \includegraphics[width=1.0\columnwidth, trim=1cm 0 1cm 0]{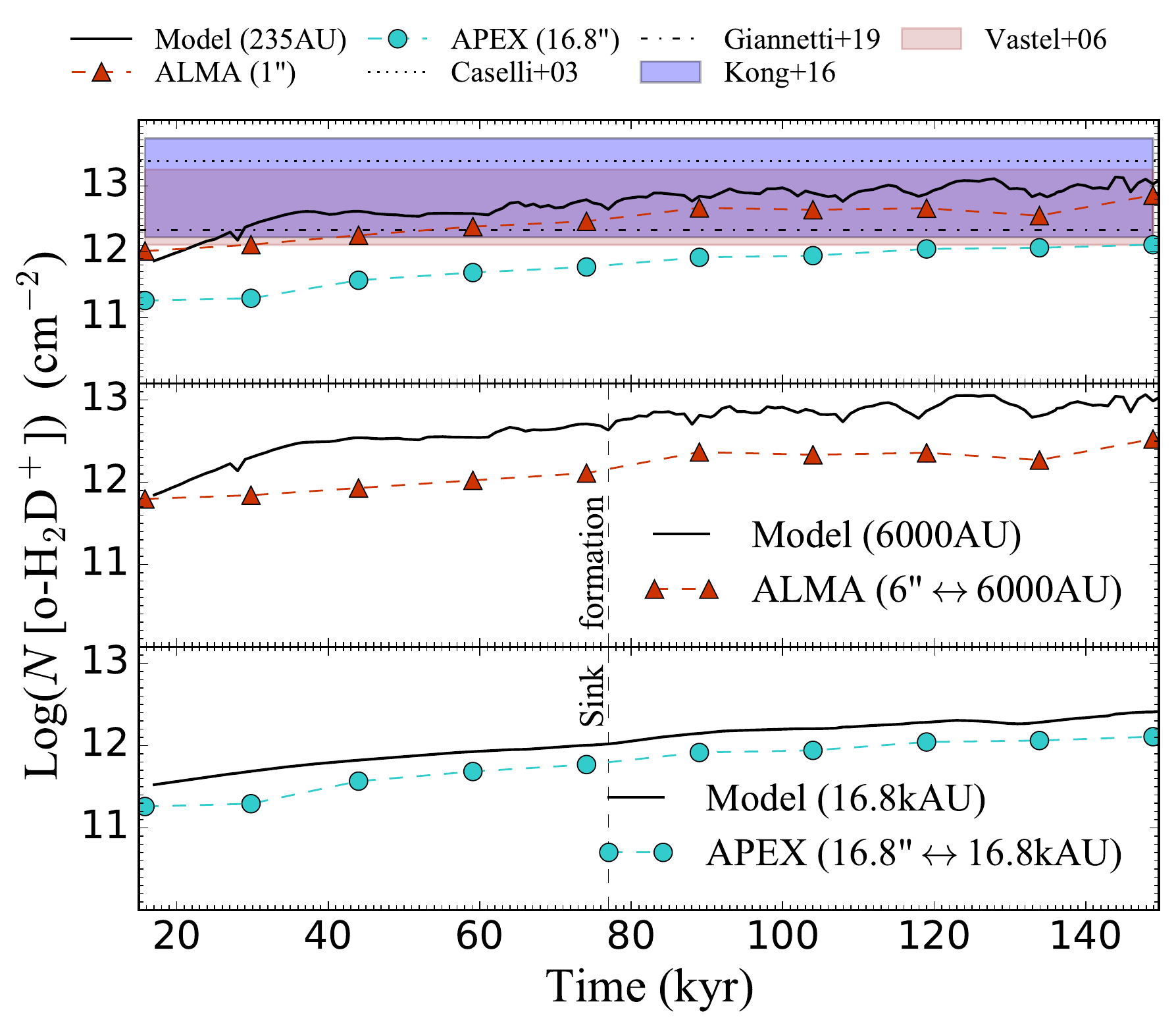}
    
    \caption[Model versus synthetic observations over collapsing time]{Model versus synthetic column densities over the collapsing  time using a source distance of 1~kpc. The upper panel show the values derived from noiseless simulation, to avoid biases based on a low signal-to-noise. The lower panel shows the simulations including thermal noise, for integration times of 8~hrs.}
    \label{fig:column_densities_information_loss}
\end{figure}

We compared our observed column densities to values reported in the literature. 
Common values of the o-H$_2$D$^+$ column density toward low- and high-mass sources are around 10$^{11}$ and 10$^{13}$~cm$^{-2}$ (see Table~\ref{tab:literature_values_of_column_density}). 
All of these values come from single-dish observations and here we compare them to our results obtained from the single-dish and interferometric simulations, aiming to understand the key differences that arise when using any of the two approaches.
Our results are presented in Fig.~\ref{fig:column_densities_information_loss}.
We show the results for the model, ALMA (only) and APEX over the core evolution, from 0.1 until 1~$t_{\rm ff}$.
The panels consider synthetic observations with (upper panel) and without (lower panel) thermal noise. 
We also report the observed values shown in Table~\ref{tab:literature_values_of_column_density} for reference.

In the upper panel, we first compared the column densities derived from the model, ALMA (only) and APEX, over their highest resolution element, i.e., 245~AU, 1" and 16.8", to see how relevant is averaging over larger areas to retrieve lower column densities.
As expected, the APEX values are lower than ALMA and the model because of its larger beam.
This means that low resolution observations tend to underestimate values by even one order of magnitude, in the single-dish case \cite[see also][]{Bovino19}.

In the second panel we show the model and ALMA values averaged over the same spatial extent, equal to 6000~AU. 
This extent covers most of the dense gas and avoids the necessity to perform multiple pointings to cover the whole image.
Performing a mosaic would require to increase the total integration time to keep a good signal-to-noise ratio of the detection, because this time is split into the number of pointings. 
In our analysis, this is only relevant when we include thermal noise in the observations.
The point of comparing over the same extent is to estimate the sensitivity of the interferometer to the real physical values.
For the noiseless case, the initial stages show an underestimate of less than one-order-of-magnitude, which gets smaller for the later stages.
When adding thermal noise to the maps, the first four timesteps (up to 60~kyr) do not follow the same trend as the model but show similar column densities along time, instead. 
Since this is not seen in the noiseless simulations, we conclude that those four points are affected by low signal-to-noise detections, because of the low central density at such stages, and that most of the emission considered within the 6" region is noise.

The third panel shows a similar comparison but for the case of APEX. 
Column densities here were averaged for both the model and APEX, over the whole APEX beam.
In this case, when averaging over the same angular scale for both the model and APEX, the differences in column densities are lower than in the case for ALMA.
This is due to the continuous spatial sampling that is achieved by single-dish telescopes, and points to the effect that the spatial filtering from interferometers has on the estimation of column densities.

\begin{table}
    \caption[Column densities of o-H$_2$D$^+$ reported in the literature]{List of column densities of o-H$_2$D$^+$ (N[o-H$_2$D$^+$]) reported in the literature and used in Fig.~\ref{fig:column_densities_information_loss}. }
    \label{tab:literature_values_of_column_density}
    \centering
    \setlength\tabcolsep{2pt}
    \begin{tabular}{lccl}
        \hline
        Telescope & FWHM (") & $\log_{10} (N$[o-H$_2$D$^+$]) (cm$^{-2}$) & Source \\
        \hline
        CSO & 22    & 13.38 & Low-mass core$^{a}$  \\
        CSO & 22    & 12.05 - 13.25 & Low-mass core$^{b}$     \\
        CSO & 22    & 12.38 - 13.61 & Low-mass core$^{c}$     \\
        JCMT & 15   & 12.28 - 12.58 & High-mass clump$^{d}$  \\
        JCMT & 15   & 12.22 - 13.72 & High-mass core$^{e}$   \\
        APEX & 16.8 & 12.33 - 12.52 & High-mass core$^{f}$ \\
        \hline
        \multicolumn{2}{l}{$^{a}$\citet{Caselli03}} & \multicolumn{1}{l}{$^{b}$\citet{Vastel06}}  & \multicolumn{1}{l}{$^{c}$\citet{Caselli08}}\\
        \multicolumn{2}{l}{$^{d}$\citet{Pillai12}} & \multicolumn{1}{l}{$^{e}$\citet{Kong16}} & \multicolumn{1}{l}{$^{f}$\citet{Giannetti19}}\\
        
    \end{tabular}
\end{table}

%

\begin{figure}
    \centering
    \includegraphics[trim=0.5cm 0.5cm 0.5cm 0.5cm, width=1.0\columnwidth]{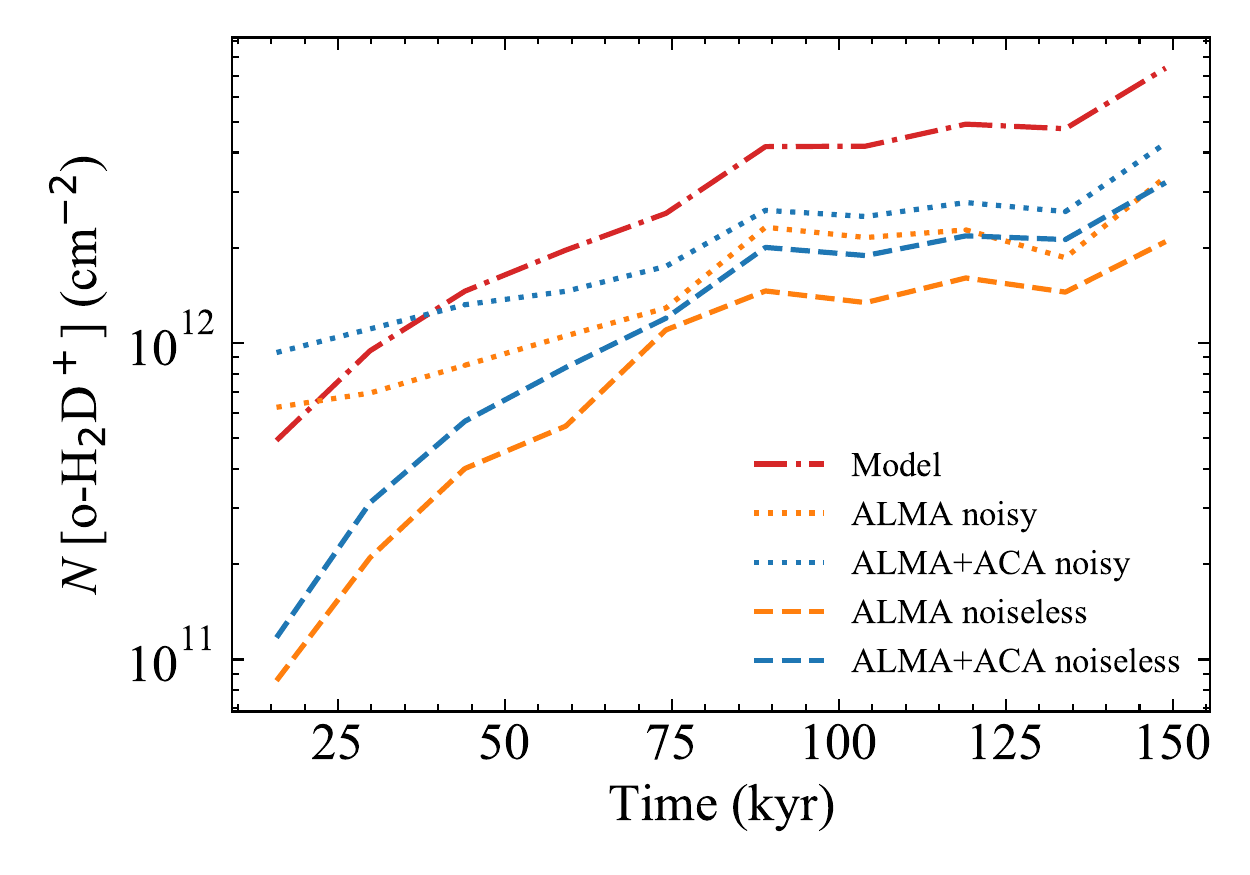}
    \caption{Column  density  evolution  over  time.   All  points  are  averaged  over  6  arcseconds,  for  the  model  (semi-dashed red), for both ALMA (orange) and ALMA+ACA (blue) in the case of including (dotted) and not including (dashed) thermal noise.}
    \label{fig:column_densities_different_configs}
\end{figure}

We also reproduced the same analysis including ACA to the ALMA noisy and noiseless simulations, to see how significant is the increase in the recovered flux when more antennas are added to the array and more points to the \textit{uv}-sampling.
From the results, shown in Fig.~\ref{fig:column_densities_different_configs}, we see that for both the noiseless and noisy simulations, the inclusion of ACA increases the retrieved column densities.
Similarly to the case in Fig.~\ref{fig:column_densities_information_loss}, the ALMA and ALMA+ACA noisy simulations show column densities larger than the model at the earliest stages. 
This overestimation occurs at the earliest stage ($\sim$16~kyr) in the case of ALMA only and until $\sim$30~kyr for ALMA+ACA, and is explained by the predominance of noise in the averaging region of 6" due to the low abundance of o-H$_2$D$^+$ at the earliest stages.

\subsection{Minimum integration time for a $5\sigma$ detection}
\label{subsec:min_int_time}
We also studied the observability of the o-H$_2$D$^+$ $J_{\rm {K_a, K_c}} = 1_{10}$-$1_{11}$ line as a function of the distance to the source and the integration time required for interferometric observations only.
We were interested in the dependence of the minimum integration time required to have a 5$\sigma$ detection on the ratio of the column density measured over the model values when the source is placed farther away. 
For this analysis we performed ALMA and ALMA+ACA simulations for the core at 74~kyr (see middle panels in Fig.~\ref{fig:mhd_snapshots} and~\ref{fig:radial_distributions_1kpc}) at different distances, from 1 up to 7~kpc. 
The results are shown in Fig.~\ref{fig:tint_vs_distance}. 
The first panel shows the minimum integration time per simulation that is required to obtain a 5$\sigma$ detection.
In order to keep the SNR constant at 5, we had to integrate longer for larger distances because the received flux decays as a function of the distance.
At a distance of 1~kpc, a SNR of 5 can be reached after $\sim$0.5-1 hrs of total integration time on source, using ALMA+ACA and ALMA, respectively. 
At a distance of 7~kpc, a SNR of 5 can be reached by at least a 6 and 7 hrs total integration time using ALMA+ACA and ALMA, respectively.
All these simulations were binned along the spectral axis by a factor of 10, from $\Delta v=0.03$~km s$^{-1}$ to 0.3~km s$^{-1}$ to decrease the noise level, following equation (\ref{eq:T_rms_single-dish}).
The SNR value considered here was obtained by measuring the peak flux within the entire data cube and dividing it by the cube rms.
The rms of each cube was obtained by deriving the mean rms over 4 rectangular regions far from the source, for a few channels far from the line emission and then averaged them all.
Simulations leading to a SNR value lower than 5 were classified as no detections.

\subsection{Quantifying the column densities information loss}
\label{subsec:Quantifying_the_underestimation_of_column_densities}
\begin{figure*}
    \centering
    \includegraphics[width=0.8\textwidth]{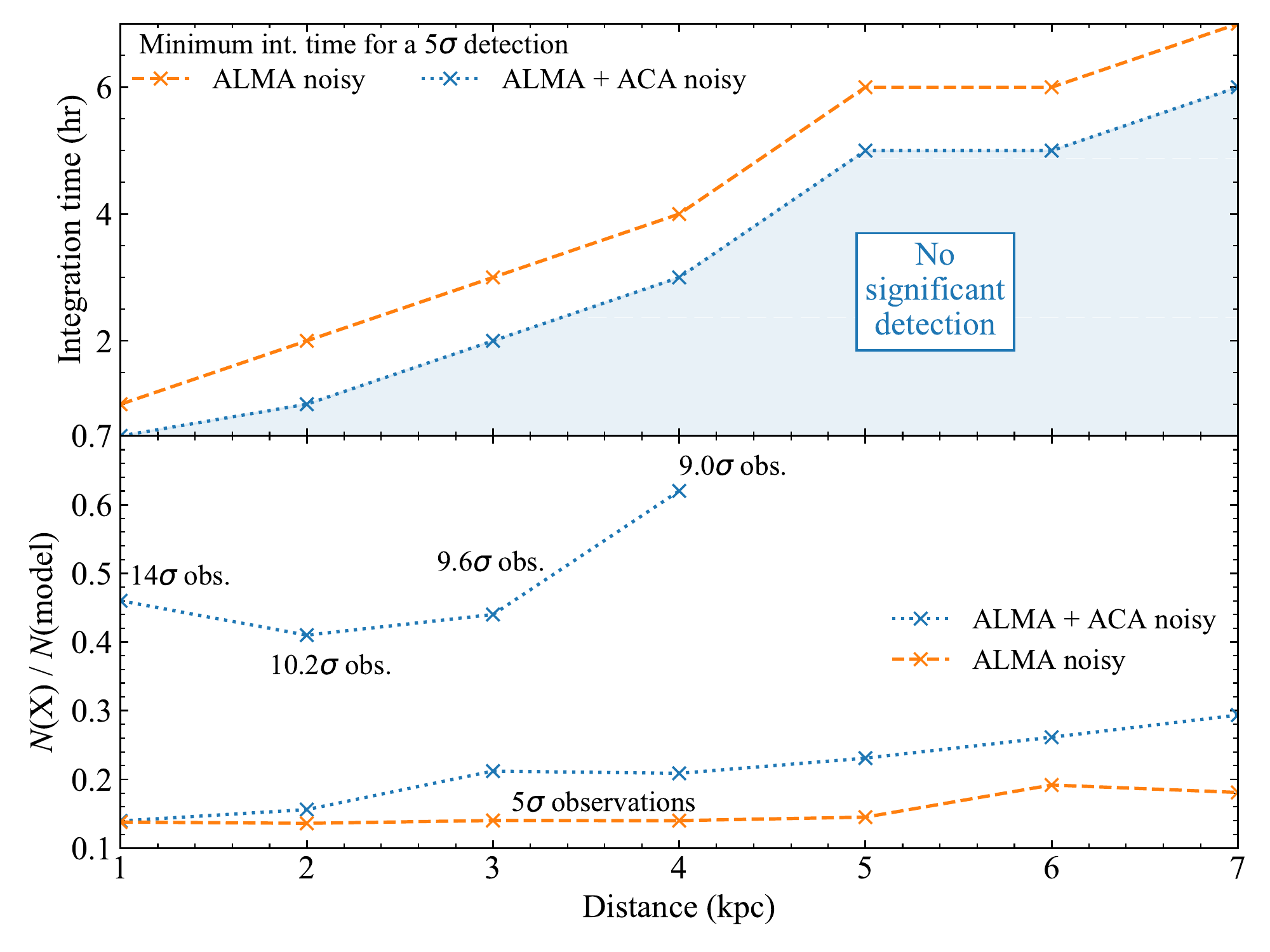}
    \caption{\textit{Top panel}: Minimum integration time required to have a 5$\sigma$ detection as a function of the distance to the source. \textit{Bottom panel}: Ratio of synthetic column densities as derived from ALMA and ALMA+ACA over the values from the model as a function of the source distance.}
    \label{fig:tint_vs_distance}
\end{figure*}

For the simulations at 5$\sigma$, we derived the synthetic column densities and then calculated the ratio of simulated values over  model values, namely, $N({\rm X}) / N({\rm model})$ for X being ALMA and ALMA+ACA outputs.
The results are shown in the bottom panel of Fig.~\ref{fig:tint_vs_distance}. 
The ALMA and ALMA+ACA 5$\sigma$ detections are quantified by the column density ratios $N_{\rm ALMA} / N_{\rm model}$ and $N_{\rm ALMA} / N_{\rm model}$. 
Similar to the analysis for noiseless maps in section~\ref{subsec:Radial_distributions} (see Figs.~\ref{fig:radial_distributions_1kpc}, \ref{fig:radial_distributions_5kpc} and \ref{fig:radial_distributions_10kpc}), at the same SNR the closest distances will show the lowest ratios.
This is because the spatial filtering becomes more effective when the source is more extended relative to the size of the beam, and then less regions from the actual source are sampled.
For these low-SNR observations, we lose around 14\% of the model column densities at 1~kpc and even 29\% at 7~kpc.
Similar to the previous analysis, the ALMA+ACA simulations recover a larger fraction of the model column densities than the pure ALMA case, because of its more complete \textit{uv}-coverage.

The column density ratios obtained by low SNR observations are rather low, since the source emission that is detected above the noise limit is a tiny fraction of the actual emission.
Therefore, a natural question to ask is how much of the real column densities we lose with a high SNR detection.
In order to understand this, we also performed ALMA+ACA simulations at the same distances but this time increasing the integration time to 10~hrs each. 
The SNR of these observations were 14, 10.2, 9.6 and 9.0 for the 1, 2, 3 and 4~kpc distances, respectively.
These simulations were also further binned in velocity to $\Delta v = 0.1$~km s$^{-1}$ to obtain high SNR values.
The 10~hrs integration time value was found by increasing the time constantly while keeping attention to the returned SNR.
It showed that for longer times, the elevation of the source was not enough to contribute to the source signal significantly and then the SNR became slightly lower.
The results for 1, 2, 3 and 4~kpc are shown in the lower panel of Fig.~\ref{fig:tint_vs_distance}.
For the high-SNR observation at 1~kpc, the model column densities are underestimated by a factor of $\sim0.54$.
For the 2~kpc case, the underestimation increases a bit to $\sim0.59$ and then decreases to $\sim0.56$ and $\sim0.48$ at 3 and 4~kpc.
The lack of a direct correlation between the signal-to-noise ratio and the column density ratio represents an interplay between the spatial filtering effect and the low recovery of source emission at lower SNR.
The former is more relevant at 1 and 2~kpc while the latter is more relevant at 3 and 4~kpc.

\section{Summary \& Conclusions}
\label{sec:Summary_and_conclusion}
In this work we have studied the observability of the molecular line emission of ortho-H$_2$D$^+ J_{\rm {K_a, K_c}} = 1_{10}$-$1_{11}$ at 372.42~GHz using the simulation of an isolated high-mass magnetized prestellar core as a synthetic source, post-processed by a radiative transfer simulation and a subsequent set of interferometric and single-dish synthetic observations. 
We studied the line emission throughout the core evolution up to one free-fall time and also for different distances at one fixed timestep.
We focused on the differences that may arise when deriving column densities from a physical perspective, i.e., integrating the number density within the simulation box and when deriving them from the observed flux.
We also focused on the differences in the column densities retrieved by a single-dish and an interferometer.
The conclusions of this work can be summarized as follows:
\begin{itemize}
    \item[-]
    Column density estimates directly depend on the size of the area over which they are averaged. 
    Then, when observing centrally concentrated density distributions, from observations performed at lower angular resolutions one will tend to have large losses due to beam dilution effects. Similar results have been also reported by \cite{Bovino19}.
    This is highly effective when comparing estimations between interferometric and single-dish observations.
    However, when comparing both approaches to the real values averaged over the respective spatial extent, single-dish telescope estimations are much closer to the model than from interferometers.
    This is the result of a spatial filtering effect that is intrinsic to interferometers, which decreases the overall source sampling as compared to single-dish telescopes, in which the sampling is continuous within the beam.
    
    \item[-]
    The combined observations of ALMA and ACA allow to reach a given signal-to-noise (SNR) in less time than with ALMA only, and also improves the column density estimation as compared to the real values because it aids in reducing the effect of spatial filtering.
    
    \item[-]
    The correlation between distance and the fraction of column density obtained by ALMA+ACA ($N_{\rm ALMA+ACA}/N_{\rm model}$) is not exactly linear. 
    Instead, it is determined by an interplay between the SNR and the spatial filtering. 
    Our results show that with a $14\sigma$ detection (SNR=14) of o-H$_2$D$^+$ at a distance of 1~kpc, the column density of the source is underestimated by a 54\% (i.e., $N_{\rm ALMA+ACA}/N_{\rm model}\sim0.46$) and even more, by a 59\% at 2~kpc, with a $10.2\sigma$ detection. 
    This indicates that a lower SNR recover less of the source emission. 
    However, at a distance of 3 and 4~kpc, this underestimation becomes 56\% ($9.6\sigma$) and 38\% ($9\sigma$), respectively.  
    Hence, we conclude that the underestimation of column densities as a function of distance is a combination of the decrease of SNR (relevant up to 2~kpc) and the effect of spatial filtering (relevant for 3 and 4~kpc).
\end{itemize}

We emphasize that the best results to estimate the physical column densities from a source will be obtained through the combination of single-dish and interferometric observations. 
To conclude, we have shown that there are two effects that contribute to the information loss of the estimated column densities, these are, beam dilution in the case of single-dish telescopes spatial filtering in the case of interferometers. 

\section*{Acknowledgements}
JZ and DRGS thank for funding via Fondecyt regular 1161247 and 1201280.
DRGS and SB also thank for funding via Conicyt Programa de Astronomia Fondo Quimal 2017 QUIMAL170001 and BASAL Centro de Astrof\'isica y Tecnolog\'ias Afines (CATA) PFB-06/2007.
SB thanks for funding through Fondecyt Iniciacion 11170268.
SB and JZ also thanks for funding through the DFG priority program “The Physics of the Interstellar Medium” (project BO 4113/1-2).
The simulations were performed with resources provided by the \emph{Kultrun Astronomy Hybrid Cluster} at the Department of Astronomy, Universidad de Concepci\'on. 
JZ thanks for kind hospitality at Hamburg Observatory, the University of Kiel and INAF in Bologna.
This research made use of NASAs Astrophysics Data System Bibliographic Services (ADS), of Astropy, a community-developed core Python package for Astronomy (\citealt{astropy2013,astropy2018}; see also http://www.astropy.org) and Matplotlib~\citep{Hunter07}.


\bibliographystyle{mnras}
\bibliography{references}

\newcommand{\noop}[1]{}
\begin{thebibliography}{}
\makeatletter
\relax
\def\mn@urlcharsother{\let\do\@makeother \do\$\do\&\do\#\do\^\do\_\do\%\do\~}
\def\mn@doi{\begingroup\mn@urlcharsother \@ifnextchar [ {\mn@doi@}
  {\mn@doi@[]}}
\def\mn@doi@[#1]#2{\def\@tempa{#1}\ifx\@tempa\@empty \href
  {http://dx.doi.org/#2} {doi:#2}\else \href {http://dx.doi.org/#2} {#1}\fi
  \endgroup}
\def\mn@eprint#1#2{\mn@eprint@#1:#2::\@nil}
\def\mn@eprint@arXiv#1{\href {http://arxiv.org/abs/#1} {{\tt arXiv:#1}}}
\def\mn@eprint@dblp#1{\href {http://dblp.uni-trier.de/rec/bibtex/#1.xml}
  {dblp:#1}}
\def\mn@eprint@#1:#2:#3:#4\@nil{\def\@tempa {#1}\def\@tempb {#2}\def\@tempc
  {#3}\ifx \@tempc \@empty \let \@tempc \@tempb \let \@tempb \@tempa \fi \ifx
  \@tempb \@empty \def\@tempb {arXiv}\fi \@ifundefined
  {mn@eprint@\@tempb}{\@tempb:\@tempc}{\expandafter \expandafter \csname
  mn@eprint@\@tempb\endcsname \expandafter{\@tempc}}}

\bibitem[\protect\citeauthoryear{{Astropy Collaboration} et~al.,}{{Astropy
  Collaboration} et~al.}{2013}]{astropy2013}
{Astropy Collaboration} et~al., 2013, \mn@doi [\aap]
  {10.1051/0004-6361/201322068}, \href
  {http://adsabs.harvard.edu/abs/2013A%26A...558A..33A} {558, A33}

\bibitem[\protect\citeauthoryear{{Bergin} \& {Tafalla}}{{Bergin} \&
  {Tafalla}}{2007}]{BerginAndTafalla07}
{Bergin} E.~A.,  {Tafalla} M.,  2007, \mn@doi [\araa]
  {10.1146/annurev.astro.45.071206.100404}, \href
  {http://adsabs.harvard.edu/abs/2007ARA%26A..45..339B} {45, 339}

\bibitem[\protect\citeauthoryear{{Bonnell}, {Bate}, {Clarke}  \&
  {Pringle}}{{Bonnell} et~al.}{1997}]{Bonnell97}
{Bonnell} I.~A.,  {Bate} M.~R.,  {Clarke} C.~J.,   {Pringle} J.~E.,  1997,
  \mn@doi [\mnras] {10.1093/mnras/285.1.201}, \href
  {http://adsabs.harvard.edu/abs/1997MNRAS.285..201B} {285, 201}

\bibitem[\protect\citeauthoryear{{Bonnell}, {Bate}, {Clarke}  \&
  {Pringle}}{{Bonnell} et~al.}{2001}]{Bonnell01}
{Bonnell} I.~A.,  {Bate} M.~R.,  {Clarke} C.~J.,   {Pringle} J.~E.,  2001,
  \mn@doi [\mnras] {10.1046/j.1365-8711.2001.04270.x}, \href
  {http://adsabs.harvard.edu/abs/2001MNRAS.323..785B} {323, 785}

\bibitem[\protect\citeauthoryear{{Bonnor}}{{Bonnor}}{1956}]{Bonnor56}
{Bonnor} W.~B.,  1956, \mn@doi [\mnras] {10.1093/mnras/116.3.351}, \href
  {http://adsabs.harvard.edu/abs/1956MNRAS.116..351B} {116, 351}

\bibitem[\protect\citeauthoryear{{Bovino}, {Ferrada-Chamorro}, {Lupi},
  {Sabatini}, {Giannetti}  \& {Schleicher}}{{Bovino} et~al.}{2019}]{Bovino19}
{Bovino} S.,  {Ferrada-Chamorro} S.,  {Lupi} A.,  {Sabatini} G.,  {Giannetti}
  A.,   {Schleicher} D.~R.~G.,  2019, \mn@doi [\apj]
  {10.3847/1538-4357/ab53e4}, \href
  {https://ui.adsabs.harvard.edu/abs/2019ApJ...887..224B} {887, 224}

\bibitem[\protect\citeauthoryear{{Brauer}, {Wolf}, {Reissl}  \&
  {Ober}}{{Brauer} et~al.}{2017}]{Brauer17}
{Brauer} R.,  {Wolf} S.,  {Reissl} S.,   {Ober} F.,  2017, \mn@doi [\aap]
  {10.1051/0004-6361/201629001}, \href
  {https://ui.adsabs.harvard.edu/abs/2017A&A...601A..90B} {601, A90}

\bibitem[\protect\citeauthoryear{{Brünken} et~al.,}{{Brünken}
  et~al.}{2014}]{Brunken14}
{Brünken} S.,  et~al., 2014, Nature, 516, 219

\bibitem[\protect\citeauthoryear{{Busquet}, {Palau}, {Estalella}, {Girart},
  {S\'anchez-Monge}, {Viti}, {Ho}  \& {Zhang}}{{Busquet}
  et~al.}{2010}]{Busquet10}
{Busquet} G.,  {Palau} A.,  {Estalella} R.,  {Girart} J.,  {S\'anchez-Monge}
  A.,  {Viti} S.,  {Ho} P.,   {Zhang} Q.,  2010, A\&A, 517, L6

\bibitem[\protect\citeauthoryear{{Caselli}}{{Caselli}}{2013}]{Caselli13}
{Caselli} P.,  2013, in {Kawabe} R.,  {Kuno} N.,   {Yamamoto} S.,  eds,
  Astronomical Society of the Pacific Conference Series Vol. 476, New Trends in
  Radio Astronomy in the ALMA Era: The 30th Anniversary of Nobeyama Radio
  Observatory. p.~169

\bibitem[\protect\citeauthoryear{{Caselli} \& {Ceccarelli}}{{Caselli} \&
  {Ceccarelli}}{2012}]{CaselliAndCeccarelli12}
{Caselli} P.,  {Ceccarelli} C.,  2012, \mn@doi [\aapr]
  {10.1007/s00159-012-0056-x}, \href
  {http://adsabs.harvard.edu/abs/2012A%26ARv..20...56C} {20, 56}

\bibitem[\protect\citeauthoryear{{Caselli}, {Walmsley}, {Tafalla}, {Dore}  \&
  {Myers}}{{Caselli} et~al.}{1999}]{Caselli99}
{Caselli} P.,  {Walmsley} C.~M.,  {Tafalla} M.,  {Dore} L.,   {Myers} P.~C.,
  1999, \mn@doi [\apjl] {10.1086/312280}, \href
  {http://adsabs.harvard.edu/abs/1999ApJ...523L.165C} {523, L165}

\bibitem[\protect\citeauthoryear{{Caselli}, {van der Tak}, {Ceccarelli}  \&
  {Bacmann}}{{Caselli} et~al.}{2003}]{Caselli03}
{Caselli} P.,  {van der Tak} F.~F.~S.,  {Ceccarelli} C.,   {Bacmann} A.,  2003,
  \mn@doi [\aap] {10.1051/0004-6361:20030526}, \href
  {http://adsabs.harvard.edu/abs/2003A%26A...403L..37C} {403, L37}

\bibitem[\protect\citeauthoryear{{Caselli}, {Vastel}, {Ceccarelli}, {van der
  Tak}, {Crapsi}  \& {Bacmann}}{{Caselli} et~al.}{2008}]{Caselli08}
{Caselli} P.,  {Vastel} C.,  {Ceccarelli} C.,  {van der Tak} F.~F.~S.,
  {Crapsi} A.,   {Bacmann} A.,  2008, \mn@doi [\aap]
  {10.1051/0004-6361:20079009}, \href
  {http://adsabs.harvard.edu/abs/2008A%26A...492..703C} {492, 703}

\bibitem[\protect\citeauthoryear{{Dalgarno} \& {Lepp}}{{Dalgarno} \&
  {Lepp}}{1984}]{Dalgarno&Lepp84}
{Dalgarno} A.,  {Lepp} S.,  1984, \mn@doi [\apjl] {10.1086/184395}, \href
  {https://ui.adsabs.harvard.edu/abs/1984ApJ...287L..47D} {287, L47}

\bibitem[\protect\citeauthoryear{{Ebert}}{{Ebert}}{1955}]{Ebert95}
{Ebert} R.,  1955, \zap, \href
  {http://adsabs.harvard.edu/abs/1955ZA.....36..222E} {36, 222}

\bibitem[\protect\citeauthoryear{{Flower}, {Pineau Des For{\^e}ts}  \&
  {Walmsley}}{{Flower} et~al.}{2006}]{Flower06}
{Flower} D.~R.,  {Pineau Des For{\^e}ts} G.,   {Walmsley} C.~M.,  2006, \mn@doi
  [\aap] {10.1051/0004-6361:20054246}, \href
  {https://ui.adsabs.harvard.edu/abs/2006A&A...449..621F} {449, 621}

\bibitem[\protect\citeauthoryear{{Friesen}, {Di Francesco}, {Bourke},
  {Caselli}, {J{\o}rgensen}, {Pineda}  \& {Wong}}{{Friesen}
  et~al.}{2014}]{Friesen14}
{Friesen} R.~K.,  {Di Francesco} J.,  {Bourke} T.~L.,  {Caselli} P.,
  {J{\o}rgensen} J.~K.,  {Pineda} J.~E.,   {Wong} M.,  2014, \mn@doi [\apj]
  {10.1088/0004-637X/797/1/27}, \href
  {https://ui.adsabs.harvard.edu/abs/2014ApJ...797...27F} {797, 27}

\bibitem[\protect\citeauthoryear{{Fryxell} et~al.,}{{Fryxell}
  et~al.}{2000}]{FLASH}
{Fryxell} B.,  et~al., 2000, \mn@doi [\apjs] {10.1086/317361}, \href
  {http://adsabs.harvard.edu/abs/2000ApJS..131..273F} {131, 273}

\bibitem[\protect\citeauthoryear{{Giannetti} et~al.,}{{Giannetti}
  et~al.}{2014}]{Giannetti14}
{Giannetti} A.,  et~al., 2014, \mn@doi [\aap] {10.1051/0004-6361/201423692},
  \href {https://ui.adsabs.harvard.edu/abs/2014A&A...570A..65G} {570, A65}

\bibitem[\protect\citeauthoryear{{Giannetti} et~al.,}{{Giannetti}
  et~al.}{2019}]{Giannetti19}
{Giannetti} A.,  et~al., 2019, \mn@doi [\aap] {10.1051/0004-6361/201834602},
  \href {https://ui.adsabs.harvard.edu/abs/2019A&A...621L...7G} {621, L7}

\bibitem[\protect\citeauthoryear{{Grassi}, {Bovino}, {Schleicher}, {Prieto},
  {Seifried}, {Simoncini}  \& {Gianturco}}{{Grassi} et~al.}{2014}]{Grassi14}
{Grassi} T.,  {Bovino} S.,  {Schleicher} D.~R.~G.,  {Prieto} J.,  {Seifried}
  D.,  {Simoncini} E.,   {Gianturco} F.~A.,  2014, \mn@doi [\mnras]
  {10.1093/mnras/stu114}, \href
  {https://ui.adsabs.harvard.edu/abs/2014MNRAS.439.2386G} {439, 2386}

\bibitem[\protect\citeauthoryear{{G{\"u}sten}, {Nyman}, {Schilke}, {Menten},
  {Cesarsky}  \& {Booth}}{{G{\"u}sten} et~al.}{2006}]{Gusten06}
{G{\"u}sten} R.,  {Nyman} L.~{\r{A}}.,  {Schilke} P.,  {Menten} K.,  {Cesarsky}
  C.,   {Booth} R.,  2006, \mn@doi [\aap] {10.1051/0004-6361:20065420}, \href
  {https://ui.adsabs.harvard.edu/abs/2006A&A...454L..13G} {454, L13}

\bibitem[\protect\citeauthoryear{{Harju} et~al.,}{{Harju}
  et~al.}{2006}]{Harju06}
{Harju} J.,  et~al., 2006, \mn@doi [\aap] {10.1051/0004-6361:20065337}, \href
  {https://ui.adsabs.harvard.edu/abs/2006A&A...454L..55H} {454, L55}

\bibitem[\protect\citeauthoryear{{Haworth}, {Glover}, {Koepferl}, {Bisbas}  \&
  {Dale}}{{Haworth} et~al.}{2018}]{Haworth18}
{Haworth} T.~J.,  {Glover} S.~C.~O.,  {Koepferl} C.~M.,  {Bisbas} T.~G.,
  {Dale} J.~E.,  2018, \mn@doi [\nar] {10.1016/j.newar.2018.06.001}, \href
  {http://adsabs.harvard.edu/abs/2018NewAR..82....1H} {82, 1}

\bibitem[\protect\citeauthoryear{{H{\"o}gbom}}{{H{\"o}gbom}}{1974}]{Hogbom74}
{H{\"o}gbom} J.~A.,  1974, \aaps, \href
  {http://adsabs.harvard.edu/abs/1974A%26AS...15..417H} {15, 417}

\bibitem[\protect\citeauthoryear{{Hugo}, {Asvany}, {Harju}  \&
  {Schlemmer}}{{Hugo} et~al.}{2007}]{Hugo07}
{Hugo} E.,  {Asvany} O.,  {Harju} J.,   {Schlemmer} S.,  2007, in Molecules in
  Space and Laboratory. p.~119

\bibitem[\protect\citeauthoryear{{Hugo}, {Asvany}  \& {Schlemmer}}{{Hugo}
  et~al.}{2009}]{Hugo09}
{Hugo} E.,  {Asvany} O.,   {Schlemmer} S.,  2009, \mn@doi [\jcp]
  {10.1063/1.3089422}, \href
  {https://ui.adsabs.harvard.edu/abs/2009JChPh.130p4302H} {130, 164302}

\bibitem[\protect\citeauthoryear{{Hunter}}{{Hunter}}{2007}]{Hunter07}
{Hunter} J.~D.,  2007, \mn@doi [Computing in Science and Engineering]
  {10.1109/MCSE.2007.55}, \href
  {https://ui.adsabs.harvard.edu/abs/2007CSE.....9...90H} {9, 90}

\bibitem[\protect\citeauthoryear{{Iguchi} et~al.,}{{Iguchi}
  et~al.}{2009}]{Iguchi09}
{Iguchi} S.,  et~al., 2009, \mn@doi [\pasj] {10.1093/pasj/61.1.1}, \href
  {https://ui.adsabs.harvard.edu/abs/2009PASJ...61....1I} {61, 1}

\bibitem[\protect\citeauthoryear{{Jorgensen}, {Belloche}  \&
  {Garrod}}{{Jorgensen} et~al.}{2020}]{Jorgensen20}
{Jorgensen} J.~K.,  {Belloche} A.,   {Garrod} R.~T.,  2020, arXiv e-prints,
  \href {https://ui.adsabs.harvard.edu/abs/2020arXiv200607071J} {p.
  arXiv:2006.07071}

\bibitem[\protect\citeauthoryear{{Klein} et~al.,}{{Klein}
  et~al.}{2014}]{FLASH+}
{Klein} T.,  et~al., 2014, IEEE Transactions on Terahertz Science and
  Technology, 4, 588

\bibitem[\protect\citeauthoryear{{Kong} et~al.,}{{Kong} et~al.}{2016}]{Kong16}
{Kong} S.,  et~al., 2016, \mn@doi [\apj] {10.3847/0004-637X/821/2/94}, \href
  {https://ui.adsabs.harvard.edu/abs/2016ApJ...821...94K} {821, 94}

\bibitem[\protect\citeauthoryear{{K{\"o}nig} et~al.,}{{K{\"o}nig}
  et~al.}{2017}]{Konig17}
{K{\"o}nig} C.,  et~al., 2017, \mn@doi [\aap] {10.1051/0004-6361/201526841},
  \href {http://adsabs.harvard.edu/abs/2017A%26A...599A.139K} {599, A139}

\bibitem[\protect\citeauthoryear{{Kraus}}{{Kraus}}{1966}]{Kraus_Radio_Astronomy}
{Kraus} J.~D.,  1966, {Radio astronomy}

\bibitem[\protect\citeauthoryear{{Körtgen}, {Bovino}, {Schleicher},
  {Giannetti}  \& {Banerjee}}{{Körtgen} et~al.}{2017}]{Kortgen17}
{Körtgen} B.,  {Bovino} S.,  {Schleicher} D.~R.,  {Giannetti} A.,   {Banerjee}
  R.,  2017, \mnras, 469, 2602

\bibitem[\protect\citeauthoryear{{Mangum} \& {Shirley}}{{Mangum} \&
  {Shirley}}{2015}]{MagnumAndShirley}
{Mangum} J.~G.,  {Shirley} Y.~L.,  2015, \mn@doi [\pasp] {10.1086/680323},
  \href {http://adsabs.harvard.edu/abs/2015PASP..127..266M} {127, 266}

\bibitem[\protect\citeauthoryear{{McKee} \& {Ostriker}}{{McKee} \&
  {Ostriker}}{2007}]{McKeeAndOstriker}
{McKee} C.~F.,  {Ostriker} E.~C.,  2007, \mn@doi [\araa]
  {10.1146/annurev.astro.45.051806.110602}, \href
  {http://adsabs.harvard.edu/abs/2007ARA%26A..45..565M} {45, 565}

\bibitem[\protect\citeauthoryear{{McKee} \& {Tan}}{{McKee} \&
  {Tan}}{2002}]{McKeeAndTan02}
{McKee} C.~F.,  {Tan} J.~C.,  2002, \mn@doi [\nat] {10.1038/416059a}, \href
  {https://ui.adsabs.harvard.edu/abs/2002Natur.416...59M} {416, 59}

\bibitem[\protect\citeauthoryear{{McKee} \& {Tan}}{{McKee} \&
  {Tan}}{2003}]{McKeeAndTan03}
{McKee} C.~F.,  {Tan} J.~C.,  2003, \mn@doi [\apj] {10.1086/346149}, \href
  {http://adsabs.harvard.edu/abs/2003ApJ...585..850M} {585, 850}

\bibitem[\protect\citeauthoryear{{Miettinen}}{{Miettinen}}{2020}]{Miettinen20}
{Miettinen} O.,  2020, \mn@doi [\aap] {10.1051/0004-6361/201936730}, \href
  {https://ui.adsabs.harvard.edu/abs/2020A&A...634A.115M} {634, A115}

\bibitem[\protect\citeauthoryear{{Motte}, {Bontemps}  \& {Louvet}}{{Motte}
  et~al.}{2018}]{MotteBontemsAndLouvet}
{Motte} F.,  {Bontemps} S.,   {Louvet} F.,  2018, \mn@doi [\araa]
  {10.1146/annurev-astro-091916-055235}, \href
  {http://adsabs.harvard.edu/abs/2018ARA%26A..56...41M} {56, 41}

\bibitem[\protect\citeauthoryear{{Parise}, {Belloche}, {Du}, {G{\"u}sten}  \&
  {Menten}}{{Parise} et~al.}{2011}]{Parise11}
{Parise} B.,  {Belloche} A.,  {Du} F.,  {G{\"u}sten} R.,   {Menten} K.~M.,
  2011, \mn@doi [\aap] {10.1051/0004-6361/201015475}, \href
  {http://adsabs.harvard.edu/abs/2011A%26A...526A..31P} {526, A31}

\bibitem[\protect\citeauthoryear{{Pillai}, {Caselli}, {Kauffmann}, {Zhang},
  {Thompson}  \& {Lis}}{{Pillai} et~al.}{2012}]{Pillai12}
{Pillai} T.,  {Caselli} P.,  {Kauffmann} J.,  {Zhang} Q.,  {Thompson} M.,
  {Lis} D.,  2012, ApJ, 751, 135

\bibitem[\protect\citeauthoryear{{Price-Whelan} et~al.,}{{Price-Whelan}
  et~al.}{2018}]{astropy2018}
{Price-Whelan} A.~M.,  et~al., 2018, \mn@doi [\aj] {10.3847/1538-3881/aabc4f},
  \href {https://ui.adsabs.harvard.edu/#abs/2018AJ....156..123T} {156, 123}

\bibitem[\protect\citeauthoryear{{Reissl}, {Wolf}  \& {Brauer}}{{Reissl}
  et~al.}{2016}]{Reissl16}
{Reissl} S.,  {Wolf} S.,   {Brauer} R.,  2016, A\&A, 593, A87

\bibitem[\protect\citeauthoryear{{Sabatini}, {Giannetti}, {Bovino}, {Brand},
  {Leurini}, {Schisano}, {Pillai}  \& {Menten}}{{Sabatini}
  et~al.}{2019}]{Sabatini19}
{Sabatini} G.,  {Giannetti} A.,  {Bovino} S.,  {Brand} J.,  {Leurini} S.,
  {Schisano} E.,  {Pillai} T.,   {Menten} K.~M.,  2019, \mn@doi [\mnras]
  {10.1093/mnras/stz2818}, \href
  {https://ui.adsabs.harvard.edu/abs/2019MNRAS.490.4489S} {490, 4489}

\bibitem[\protect\citeauthoryear{{Sabatini} et~al.,}{{Sabatini}
  et~al.}{subm}]{SabatiniSUB}
{Sabatini} G.,  et~al., 2020, subm., \aap

\bibitem[\protect\citeauthoryear{{Schöier}, {van de Tak}, {van Dishoeck}  \&
  {Black}}{{Schöier} et~al.}{2005}]{Schoier05}
{Schöier} F.,  {van de Tak} F.,  {van Dishoeck} E.,   {Black} J.~H.,  2005,
  A\&A, 432, 369

\bibitem[\protect\citeauthoryear{{Shu}, {Adams}  \& {Lizano}}{{Shu}
  et~al.}{1987}]{Shu87}
{Shu} F.~H.,  {Adams} F.~C.,   {Lizano} S.,  1987, \mn@doi [\araa]
  {10.1146/annurev.aa.25.090187.000323}, \href
  {http://adsabs.harvard.edu/abs/1987ARA%26A..25...23S} {25, 23}

\bibitem[\protect\citeauthoryear{{Sipilä}, {Caselli}  \& {Harju}}{{Sipilä}
  et~al.}{2013}]{Sipila13}
{Sipilä} O.,  {Caselli} P.,   {Harju} J.,  2013, A\&A, 554, A92

\bibitem[\protect\citeauthoryear{{Swift}}{{Swift}}{2009}]{Swift09}
{Swift} J.~J.,  2009, \mn@doi [\apj] {10.1088/0004-637X/705/2/1456}, \href
  {https://ui.adsabs.harvard.edu/abs/2009ApJ...705.1456S} {705, 1456}

\bibitem[\protect\citeauthoryear{{Tafalla}, {Myers}, {Caselli}, {Walmsley}  \&
  {Comito}}{{Tafalla} et~al.}{2002}]{Tafalla02}
{Tafalla} M.,  {Myers} P.~C.,  {Caselli} P.,  {Walmsley} C.~M.,   {Comito} C.,
  2002, \mn@doi [\apj] {10.1086/339321}, \href
  {http://adsabs.harvard.edu/abs/2002ApJ...569..815T} {569, 815}

\bibitem[\protect\citeauthoryear{{Vastel}, {Phillips}  \& {Yoshida}}{{Vastel}
  et~al.}{2004}]{Vastel04}
{Vastel} C.,  {Phillips} T.~G.,   {Yoshida} H.,  2004, \mn@doi [\apjl]
  {10.1086/421265}, \href {http://adsabs.harvard.edu/abs/2004ApJ...606L.127V}
  {606, L127}

\bibitem[\protect\citeauthoryear{{Vastel}, {Caselli}, {Ceccarelli}, T.,
  {Wiedner}, {Peng}, {Houde}  \& {Dominik}}{{Vastel} et~al.}{2006}]{Vastel06}
{Vastel} C.,  {Caselli} P.,  {Ceccarelli} C.,  T. P.,  {Wiedner} M.,  {Peng}
  R.,  {Houde} M.,   {Dominik} C.,  2006, ApJ, 645, 1198

\bibitem[\protect\citeauthoryear{{Walmsley}, {Flower}  \& {Pineau de
  Forêts}}{{Walmsley} et~al.}{2004}]{Walmsley04}
{Walmsley} C.,  {Flower} D.,   {Pineau de Forêts} G.,  2004, A\&A, 418, 1035

\bibitem[\protect\citeauthoryear{{Wootten} \& {Thompson}}{{Wootten} \&
  {Thompson}}{2009}]{Wootten09}
{Wootten} A.,  {Thompson} A.~R.,  2009, \mn@doi [IEEE Proceedings]
  {10.1109/JPROC.2009.2020572}, \href
  {https://ui.adsabs.harvard.edu/abs/2009IEEEP..97.1463W} {97, 1463}

\bibitem[\protect\citeauthoryear{{Zinnecker} \& {Yorke}}{{Zinnecker} \&
  {Yorke}}{2007}]{ZinneckerAndYorke07}
{Zinnecker} H.,  {Yorke} H.~W.,  2007, \mn@doi [\araa]
  {10.1146/annurev.astro.44.051905.092549}, \href
  {http://adsabs.harvard.edu/abs/2007ARA%26A..45..481Z} {45, 481}

\bibitem[\protect\citeauthoryear{{van der Tak}, {Caselli}  \&
  {Ceccarelli}}{{van der Tak} et~al.}{2005}]{van_der_Tak05}
{van der Tak} F.~F.~S.,  {Caselli} P.,   {Ceccarelli} C.,  2005, \mn@doi [\aap]
  {10.1051/0004-6361:20052792}, \href
  {http://adsabs.harvard.edu/abs/2005A%26A...439..195V} {439, 195}

\makeatother
\end{thebibliography}

\bsp	
\label{lastpage}
\end{document}